\documentclass[11pt]{article}

\usepackage{mheck}
\usepackage{tikz}
\usepackage{hyperref}%\usepackage{footmisc}
\usepackage{tikz,textcomp}
\usepackage[normalem]{ulem}

\newcommand{\rem}[1]{} % Auskommentieren von Absaetzen
\def\C{\mathbb{C}}
\def\Z{\mathbb{Z}}
\def\Q{\mathbb{Q}}
\def\R{\mathbb{R}}

\def\P{\mathbb{P}}

\def\id{\operatorname{id}}

\def\Hirz[#1]{\mathbbm{F}_{#1}}
\def\o[#1]{\overline{#1}}

\makeatletter
\newcommand\xleftrightarrow[2][]{%
  \ext@arrow 9999{\longleftrightarrowfill@}{#1}{#2}}
\newcommand\longleftrightarrowfill@{%
  \arrowfill@\leftarrow\relbar\rightarrow}
\makeatother

\newcommand{\tors}{\mbox{tors}}
\newcommand{\pic}{\mbox{Pic}}
\newcommand{\ord}{\mbox{ord}}

\abstract{We study six dimensional supergravity theories with superconformal sectors (SCFTs). Instances of such theories can be engineered using type IIB strings, or more generally F-Theory, which translates field theoretic constraints to geometry. Specifically, we study the fate of the discrete 2-form global symmetries of the SCFT sectors. For both $(2,0)$ and $(1,0)$ theories we show that whenever the charge lattice of the SCFT sectors is non-primitively embedded into the charge lattice of the supergravity theory, there is a subgroup of these 2-form symmetries that remains unbroken by BPS strings. By the absence of global symmetries in quantum gravity, this subgroup much be gauged. Using the embedding of the charge lattices also allows us to determine how the gauged 2-form symmetry embeds into the 2-form global symmetries of the SCFT sectors, and we present several concrete examples, as well as some general observations. As an alternative derivation, we recover our results for a large class of models from a dual perspective upon reduction to five dimensions.} 

\begin{document}

\begin{center}
\vspace*{ 2.0cm}
{\Large {\bf Gauged 2-form Symmetries in 6D SCFTs Coupled to Gravity}}\\[12pt]
\vspace{-0.1cm}
\bigskip
\bigskip 
{
{{Andreas P. Braun}$^{\,\text{a }}$}, {{Magdalena Larfors}$^{\,\text{b }}$} and  {{Paul-Konstantin~Oehlmann}$^{\,\text{b }}$} 
\bigskip }\\[3pt]
\vspace{0.cm}
{\it 
 ${}^{\text{a}}$Department of Mathematical Sciences, Durham University,  Upper Mountjoy Campus, Stockton~Rd, Durham DH1 3LE, UK  \\
  ${}^{\text{b}}$Department of Physics and Astronomy, Uppsala University,  L\"agerhyddsv. 1, \\
SE-751 20 Uppsala, Sweden    
}
\\[2.0cm]
\end{center}

\maketitle

\tableofcontents

\section{Introduction}
Superconformal field theories (SCFTs) occupy a distinguished position among quantum field theories and provide us with a coarse classification by mapping quantum field theories to their fixed points under RG flow. Six dimensional SCFTs are of particular interest as this is the highest dimension in which such theories can occur \cite{Nahm:1977tg}. Via compactification, 6D SCFTs give rise to many SCFTs in lower dimensions. This approach was initiated in \cite{Ganor:1996pc} and has seen substantial recent progress particularly for four- and five-dimensional theories \cite{Gaiotto:2015usa,Razamat:2016dpl,Bah:2017gph,Kim:2017toz,Kim:2018bpg,Kim:2018lfo,Razamat:2018gro,Ohmori:2018ona,Razamat:2019ukg,Apruzzi:2019vpe,Apruzzi:2019opn,Apruzzi:2019enx, Apruzzi:2019kgb,Bhardwaj:2018yhy,Bhardwaj:2018vuu,Bhardwaj:2019jtr,Bhardwaj:2019fzv,Bhardwaj:2019xeg,1869947}.

SCFTs in six dimensions are necessarily strongly coupled and contain tensionless strings in their spectrum, which makes them particularly hard to tackle via conventional field theory means. String theory has played key role to construct \cite{Witten:1995ex,Seiberg:1996vs} and classify \cite{Heckman:2013pva,Heckman:2015bfa} those theories in geometric terms using F-Theory on non-compact singular Calabi-Yau varieties (also see the review \cite{Heckman:2018jxk}). Intuitively, these geometries must be chosen not to contain any length scales in order not to spoil conformality, and in particular need to be non-compact so that gravity is decoupled. The tensionless strings that are a hallmark of 6D SCFTs can be understood in this setting from branes wrapped on appropriate collapsed cycles. 

These SCFTs admit a wide range of global symmetries which typically can be obtained from the underlying F-Theory geometry. Besides global symmetries acting on local operators, there are also higher-form symmetries \cite{Gaiotto:2014kfa} that act on non-local operators such as the surface defects in 6D SCFTs \cite{DelZotto:2015isa}. A quintessential example of such higher-form symmetries arise in gauge theories with only adjoint matter, which have a discrete centre 1-form symmetry that acts on Wilson line operators. The BPS strings present in 6D theories likewise naturally couple to 2-form fields, which gives rise to `Wilson surfaces' that can transform under discrete 2-form symmetries. These lead to p-form symmetries with p $\leq 2$ upon compactification \cite{Albertini:2020mdx,DelZotto:2020esg,Bhardwaj:2020phs,Gukov:2020btk}. Furthermore, p-form symmetries for different p can mix under group multiplication, which gives rise to 2-groups \cite{Cordova:2018cvg,Delcamp:2018wlb,Benini:2018reh,Cordova:2020tij,DelZotto:2020sop,Apruzzi:2021vcu}.

String theory and its connection to geometry has equally prominently been used in recent years to explore consistency conditions for theories of quantum gravity. The central idea is that (compactifications of) string theory automatically engineer such theories, and to contrast this with purely field theoretic constructions. This is called the {\it swampland program} \cite{Vafa:2005ui} \footnote{See \cite{Palti:2019pca} for an introduction, review and a more extensive list of references.}, and has been successfully used to find or conjecture new and often subtle consistency conditions that need to be obeyed for field theories to have a UV completion that includes quantum gravity. Theories that obey all such conditions are said to be in the landscape, while the others are in the swampland. A classic example is the absence of global symmetries \cite{Misner:1957mt,Banks:2010zn}, a constraint that includes higher form global symmetries \cite{Harlow:2018tng}. On the one hand, this can be argued for from the effective field theory point of view using black holes, an argument that has recently been sharpened using the AdS/CFT correspondence \cite{Harlow:2018jwu,Harlow:2018tng}. On the other hand, it is a general feature of (perturbative) string theory that global symmetries on the worldsheet become gauged in space-time \cite{Banks:1988yz}.

The swampland program strives to identify the precise location of the boundary between quantum field theories that can be consistently coupled to gravity and those that cannot. To simplify this problem, one can instead study the boundary between the landscape and the swampland for the case of conformal field theories. For theories that are engineered geometrically, the RG flow has a geometric counterpart in degenerations of the geometry, and the relationship between conformal theories and non-conformal ones is analogous  to how singularities provide seeds that produce non-trivial topology upon resolution. The question about which SCFT sectors can coexist with gravity then becomes a question about certain maximal degenerations in geometry. In this work we start addressing this question for superconformal field theories in dimension six, where we can use the strategy to study compactifications of F-Theory on compact elliptically fibered Calabi-Yau threefolds. Such an approach has been previously taken in \cite{DelZotto:2014fia}, which analysed how the unimodular charge lattice required in 6D supergravity \cite{Seiberg:2011dr} is found by combining the non-unimodular charge lattices arising from the singular geometry, and those of the SCFT sectors into the unimodular lattice of the base geometry. However, it cannot be possible to couple every single 6D SCFT to gravity. While there are infinitely many 6D SCFTs with charge lattices of arbitrarily high ranks, there are only finitely many compact bases for elliptic Calabi-Yau threefolds \cite{1993alg.geom..5002G,Grassi:1991ws}. Unfortunately the list of such bases is unknown, as the proof of finiteness is not constructive. Similarly, \cite{Tarazi:2021duw} recently argued for the finiteness of massless modes in 6D $(1,0)$ supergravities from the field theory point of view. 

Whereas the SCFT sectors in a specific model can have global symmetries, such symmetries need to be gauged or broken when coupling to gravity. This can be easily understood for 
ordinary (0-form) flavour symmetries, as any flavour brane supporting them is necessarily compact in a model coupled to gravity. It is only when zooming in to any of the non-compact, local patches that give rise to one of the SCFT sectors that such branes become effectively non-compact, the gauge coupling effectively goes to zero and thus the gauge symmetry appears to be global. The absence of 2-form global symmetries $G_S$ likewise follows from the self-duality of the charge lattice of supergravity theories. The breaking of the 2-form global symmetries $G_S$ can furthermore be understood from the defects that have a non-trivial transformation behaviour becoming BPS strings of finite mass. It is not true, however, that all of the defects present in SCFT sectors necessarily persist as finite mass objects when coupling to supergravity, and thus some 2-form symmetries may remain unbroken.  

As we shall see, this happens precisely when the charge lattice $\Lambda_S$ of the SCFT sectors does not embed primitively into the charge lattice $\Lambda_B$ of the supergravity model. We find that both for $(2,0)$ and $(1,0)$ theories, the subgroup $G$ of $G_S$ which remains unbroken by BPS strings is given by 
\begin{equation}\label{eq:master_result}
G = \tors (\Lambda_B/\Lambda_S) \, .
\end{equation}
In the absence of other effects that can break $G_S$, it then follows that $G$ must be gauged.  This claim can be further substantiated by using string dualities, in particular a form of fibre-base duality. When going to five dimensions on $\mathbb{S}^1$, the 2-form symmetries in question give rise to 1-form symmetries. For models that permit an appropriate second elliptic fibration, we can then employ results on the gauging of 1-form symmetries in models with torsional Mordell-Weil groups to argue for $G$ as a gauged 2-form symmetry.

This paper is structured as follows: In Section~\ref{sect:6DSCFT_2form} we give a quick recap of $(2,0)$ and $(1,0)$ 6D SCFTs and in particular their 2-form symmetries. This section mainly serves to fix notation and to rederive that we can understand discrete 2-form symmetries in 6D as a finite subgroup of $U(1)^r$ left unbroken by BPS strings. 

In Section~\ref{sect:20andgravity} we examine how $(2,0)$ SCFTs are coupled to gravity by considering IIB compactification on K3 surfaces. The geometric condition for which SCFT subsectors that can be present can be simply phrased in terms of embeddings of the charge lattices of $(2,0)$ SCFTs into the even unimodular lattice $\Lambda_{5,21}$ of signature $(5,21)$. For a large class of such models, it is possible to employ fibre-base duality to give an alternative derivation of the gauged subgroup $G$ of $G_S$. We also present a number of explicit examples where this is the case and in which \eqref{eq:master_result} can be worked out by various techniques.  

In Section~\ref{sect:10andgravity} we then move to the broader class of $(1,0)$ theories and their F-Theory realisation. The geometric condition for which SCFTs can be coupled to gravity is now that the associated singularities can coexist in a compact elliptically fibered Calabi-Yau threefold $X_o$ with base $B_o$. Such theories can be constructed by blowing down curves spanning a lattice $\Lambda_S$ in the base $B$ of a smooth threefold $X$. Contrary to the case of $(2,0)$ theories, $\Lambda_B$ is hence no longer unique, but depends on $B$. We investigate the fate of 2-form symmetries by focussing on cases where both the base $B$ of $X$ and $B_o$ of $X_o$ are toric surfaces, and furthermore $B_o$ is such that no more curves can be collapsed. For such `extremal' cases, we find a simple condition that determines if part of the group of global 2-form symmetries becomes gauged when coupling the SCFT sector to gravity. Again, our results can be given an alternative derivation in cases where a dual description exists. Furthermore, we also discuss the gauging of 2-form symmetries for Little String Theories as intermediate cases between SCFTs and supergravity. 

Finally, we give conclusions and point out some further connections and possible approaches for future research in Section~\ref{sect:outro}. Technical details about lattices and some background on how to construct elliptic Calabi-Yau threefolds over toric bases are discussed in the appendices.  

%%%%%%%%%%%%%%%%%%%%%%%%%%%%%%%%%%%%%%%%%%%%%%%%%%%%%
%%%%%%%%%%%%%%%%%%%%%%%%%%%%%%%%%%%%%%%%%%%%%%%%%%%%%

\section{6D SCFTs and 2-form symmetries}\label{sect:6DSCFT_2form}

6D SCFT are strongly coupled theories that contain tensionless strings among their  degrees of freedom. Scale invariance alone implies that these theories are decoupled from gravity. In addition, these theories can only exist at strong coupling, as also the 6D gauge coupling admits a mass scale and goes to infinity in the IR \cite{Witten:1995ex,Seiberg:1996vs}. 
Using IIB string theory or F-Theory, 6D SCFTs are engineered by compactifications on (possibly) singular two dimensional spaces of the form  
   \begin{align}
   B_{\Gamma} =\mathbb{C}^2/\Gamma \, \quad \text{ with } \Gamma \in U(2) \, ,
   \end{align}
where the non-compactness of $B_{\Gamma}$  ensures that gravity is decoupled, and hence eliminates all scale dependence. The SCFT can be understood in terms of its tensor branch where a field theory description is available. In geometric terms this amounts to a blow-up of the base $B_{\Gamma}$ to a smooth geometry $B$. In general we are interested in theories with gauge or flavor groups, which may be engineered by D7 branes that wrap, respectively, compact and non-compact directions. F-Theory allows to systematically describe such features in terms of an elliptic three-fold
   \begin{align}
   \label{eq:elliptic3fold}
   \begin{array}{ccl}
   T &\rightarrow & X \\
   & & \downarrow \pi \\
    & & B 
   \end{array}
   \end{align}
where D7 branes are encoded by the singularity structure of the fibre torus over complex curves $C$ in $B$. This construction engineers D7 flavor branes and keeps automatically track of gauge algebra factors over curves $C \subset B$.

The curves $C$ may also be wrapped by D3 branes, thereby giving massive BPS strings in the 6D spacetime.  The tension of these strings is determined by the volume of the curve $C$, which in turn is controlled by a K\"ahler parameter. Tuning the K\"ahler parameter, one may shrink the curve to zero volume, thus rendering the strings to be tensionless. At this point, the mass scale is removed from the theory, and an SCFT description emerges.  A defining property of the SCFT is the charges of the tensionless strings, which is  determined by the intersection matrix of the base curves $C$ 
\begin{align}
C_i \cdot C_j =\Omega_{ij} \, .
\end{align}
This intersection matrix is also important, as it appears in the 6d anomaly cancellation terms, via the GSSW mechanism \cite{Green:1984bx,Sagnotti:1992qw}. There the tensor fields can shift and cancel one loop anomalies and their coupling matrix with respect to to the $I_8$ anomaly 8-form polynomial is specified via $\Omega$.  

More generally,  the structure of $\Omega$ allows to distinguish the three  types of theories that can be engineered in this fashion: 
\begin{enumerate}
\item {\bf SCFTs} if $\Omega$ is negative definite; in this case all compact curves $C_i$ may be shrunken to a point $B_\Gamma$ \cite{10.1007/BFb0097582,Mumford:1961vq}, 
\item  {\bf Little string theory} (LSTs) \cite{Ganor:1996mu} if $\Omega$ is negative semi definite such that the $C_i$ can be shrunk up to a curve of self-intersection $0$ that sets the little string tension, 
\item {\bf 6D Gravity} if  det$(\Omega)=1$ and self-dual, the base is compact and its volume sets the 6D Planck scale.
\end{enumerate}
As we will see below, the SCFTs take a center stage in our discussion; they can be glued to LSTs and also naturally sit inside a generic 6D gravity theories. 
The amount of supersymmetry is the main qualifier for the complexity of 6D theories that we want to consider here. In the following two subsections we will first review theories with maximal and minimal SUSY in 6D.

\subsection{$(2,0)$ theories}
In the absence of gravity, theories with $(2,0)$ SUSY only admit tensor representations. To preserve this amount of SUSY requires the base to be Calabi-Yau itself, hence $\Gamma \in SU(2)$. This type of singularities admit an $ADE$ classification, reflected in the intersection structure of the resolution curves $C$ of self-intersection $C^2=-2$, which hence have trivial normal bundles as required. The intersection form of the curves $C_i$  coincides precisely with the negative of the ADE Cartan matrix
\begin{align}
\Omega_{(2,0)SCFT} =  -G_{ADE} \, .
\end{align} 
Due to the large amount of SUSY there are no matter or vector multiplets. The tensor multiplets contain self-dual 2-form fields, which in the F-Theory description arise from the reduction of $C_4$ along the curves $C_i$ given above. These couple to massive BPS strings in 6D, that become massless when going to the singular point.  

LSTs are engineered in a similar fashion, but the geometry contains an additional curve $C_0$ with zero eigenvalue under the intersection form $\Omega$. This curve can not be shrunk and sources a non-dynamical tensor whose scalar component sets the little string scale in the IR. From the point of the intersection form $\Omega$, the curve $C_0$ plays exactly the role of the unique affine node to the ADE configuration. Hence the intersection form of LSTs are given by the affine Cartan matrix $\Omega= -\hat{G}$.   

\subsection{$(1,0)$ theories}
(1,0) theories have a very rich structure, as they allow for vector and hypermultiplets representations and thus non-trivial gauge and flavor groups. From the IIB perspective those vectors live on the worldvolume of spacetime filling D7 branes, which break half of the SUSY and wrap curves inside of the complex two-dimensional compactification space $B_2$. Since the D7 branes backreact on the IIB axio-dilaton, this results in a non-constant profile, which is best captured via the geometry of the elliptic threefold \eqref{eq:elliptic3fold}
of F-Theory.
  The power of this geometric construction therefore allows a geometric classification of those theories via F-Theory \cite{Heckman:2013pva,Heckman:2015bfa} (see \cite{Heckman:2018jxk} for a review and more references). When resolved, the base of a (1,0) SCFT is still built from trees of $\mathbb{P}^1$ of negative self intersection, as dictated by shrinkability \cite{Mumford:1961vq,10.1007/BFb0097582}. However the overall structure is not confined to be of ADE type anymore, nor are the curves $C$ required to have a trivial normal bundle i.e. $C^2 = -2$. This affects the elliptic fibration to be non-trivial by enforcing an ADE singularity in the elliptic fibre in case the self intersection $C^2 = -n$ is below $n>2$. These single curve theories constitute the simplest class, these rank one 6D SCFTs are called  {\it non-Higgsable cluster} (NHC) \cite{Morrison:2012np} and have the following minimal gauge group
\begin{align}
\label{eq:SingleNHCs}
\begin{array}{|c||c|c|c|c|c|c|c|c|c|} \hline
n & 1 & 2  & 3 & 4 & 5 & 6 & 7 &  8 & 12 \\ \hline
\mathfrak{g}& - & - & \mathfrak{su}_3 & \mathfrak{so}_8 & \mathfrak{f}_4 & \mathfrak{e}_6 & \mathfrak{e}_7 & \mathfrak{e}_7 & \mathfrak{e}_8 \\ \hline
\end{array}\, .
\end{align}
Among the above theories, only one admit matter representations (the $\frac12 \mathbf{56}$ of $\mathfrak{e}_7$ on the $-7$ curve). Such matter representations transform under a global flavor symmetry, which constitutes another piece of the defining 6d SCFT data, but is then trivial for most of the above NHCs. 

By enhancing the flavor symmetry further (Higgsable) SCFTs can be constructed from NHCs which enhances the minimal gauge group. Such enhancements can be conveniently be obtained by tuning certain polynomial deformations in the geometric Weierstrass model that determines the elliptic fibration \eqref{eq:elliptic3fold}. 
 As an example,  the $\mathfrak{su}_3$ over the -3 curve can be enhanced to an $\mathfrak{g}_2$ with a massless hypermultiplet in the  $\mathbf{7}$ that transforms in the $\mathfrak{su}_2$ flavor representation. Moreover, the minimal gauge symmetry over some curve can be enhanced by adjoining another $-n$ curve. E.g. the $\mathfrak{g}_2$ enhancement over the $-3$ curve can also be achieved by adding a $-2$ curve over which the aforementioned $\mathfrak{su}_2$ flavor symmetry is gauged, completed with the respective massless matter multiplets. Such a configuration then naturally leads to higher rank multi-node NHCs.

The flavor symmetries, or in the IIB context non-compact D7 flavor branes and their intersections in a smooth point, naturally lead to  {\it superconformal matter} theories \cite{DelZotto:2014hpa}, another building block of SCFTs. Such theories are generalisations of bifundamental matter which however is not perturbatively realized. This is consistently encoded in the geometry of the elliptic threefold $X$, as certain non-minimal singularities in the fibre that are avoided upon blow-up(s) of the intersection point in $B_2$.  A typical example is the intersection of an $\mathfrak{e}_8$ and $\mathfrak{su}_1$ brane that leads to a non-minimal singularity at their intersection point,  which  can be avoided by inserting a single $-1$ curve at that point. In fact this is the first theory shown in \eqref{eq:SingleNHCs} which hence admits a generic $\mathfrak{e}_8$ flavor algebra. Various  $\hat{G}_1 \times \hat{G}_2$ collisions are possible and have been studied e.g. in \cite{DelZotto:2014hpa} which further enriches the $\mathcal{N}=(1,0)$ structure\footnote{6D SCFTs with non-trivial 1-form symmetries have been studied e.g. in \cite{Ohmori:2018ona,Dierigl:2020myk}.}. However, as we will review in the next sections, such theories do not give rise to 2-form symmetries as they are no singularities of the base associated with them. 

Similar as for the (2,0) theories, there is the straightforward generalisation of 6D SCFTs to an little string theory by adding an additional rational curve such that there is a zero-eigenvalue in $\Omega$. Since the base is not necessarily of ADE type anymore, the little string curve $C_0$ is not directly identified as the affine extension anymore, which makes the classification of LSTs in terms of 6D SCFTs much richer \cite{Bhardwaj:2015oru,Bhardwaj:2019hhd} than their (2,0) counterparts.

%%%%%%%%%%%%%%%

\subsection{2-form symmetries in 6D SCFTs }
\label{sect:2form_in_6D}

In 6D, SCFTs and LSTs can have global 2-form symmetries  \cite{Gaiotto:2014kfa,Garcia-Etxebarria:2019cnb,Bhardwaj:2019fzv,Bhardwaj:2020phs,Morrison:2020ool}, with the  2-form symmetry  group being the same as the 'defect group' discussed in \cite{DelZotto:2015isa}. In contrast, 6D gravity theories do not admit global symmetries of any sort. These 2-form symmetries can be found  by working out the Smith normal form \cite{Bhardwaj:2020phs} of the inner form on $\Lambda_S = H^{2}(B,\mathbb{Z})$. In this section we show that the global 2-form symmetry of 6D SCFTs and LSTs can also be written in terms of the discriminant group of $\Lambda_S$. This discussion is accompanied by Appendix \ref{app:unimodularity}, where we review basic definitions and properties of lattices and give further comments on the equivalence of our presentation and that of Ref.~\cite{Bhardwaj:2020phs}.

A higher form symmetry is a global symmetry $G$ that acts on extended objects \cite{Gaiotto:2014kfa}. To construct it, recall that  for a 0-form (ordinary) global 
symmetry in $d$-dimensions, the conserved charge is 
\begin{equation}
Q(M^{d-1}) = \int_{M^{d-1}} \ast j \, ,
\end{equation}
where $j$ is the 1-form current. We can think of ${M^{d-1}}$ as a hypersurface in space-time (i.e. a space-like slice). For continuous symmetries, we can exponentiate $Q$ to find the symmetry operator, but we can think
more generally of associating an operator $U_g(M^{d-1})$ to $M^{d-1}$. With this more abstract perspective, the group structure is reflected by the multiplication law \cite{Gaiotto:2014kfa}
\be
U_g(M^{d-1}) U_{g'}(M^{d-1}) = U_{g''} (M^{d-1}) \, ,
\ee
where $g'' = g' g \in G$. For ordinary Abelian (non-Abelian) symmetries, this multiplication is commutative (non-commutative).    
$U_g(M^{d-1})$ is topological in the sense that it only changes when it crosses 
a local operator $V(P)$ located at $P$. To find the transformation of this operator we can use an $S^{d-1}$ surrounding $P$ and write 
\begin{equation}
U_g(S^{d-1}) V(P) = R(g) V(P) \, ,
\end{equation}
where $R(g)$ is some representation of $G$.

We can repeat the same logic for higher-dimensional operators $V(C_p)$ of dimension $p$. $M^{d-1}$ is replaced by $M^{d-1-p}$ and $S^{d-1}$ is replaced 
by a sphere $S^{d-1-p}$ linking the operator $V(C_p)$.  The multiplication law 
\be
U_g(M^{d-1-p}) U_{g'}(M^{d-1-p}) = U_{g''} (M^{d-1-p})
\ee  
for codimension $p+1$ submanifolds $M^{d-1-p}$ is, however, necessarily commutative, and so  $G$ must  be Abelian for all higher form symmetries. In particular, for 2-forms in 6D, the charged operators $V(C_2)$ are two-dimensional and may be linked by an $S^3$ leading to
\begin{equation}\label{eq:2fsym6Drep}
U_g(S^{3}) V(C_2) = R(g) V(C_2) \; .
\end{equation}

Consider now a 6D SCFT realized in F-Theory on an elliptically fibered Calabi--Yau threefold $X$ that has a non-compact base $B$ with $\Lambda_S = H_2(B,\Z)$. 
As mentioned in the beginning of this section, $D3$ branes wrapping the compact, contractable curves associated to $\Lambda_S$ give rise to tensionless strings in the 6D theory, which are charged under the 2-form gauge fields $B_i$ of the tensor multiplets. Consequently, $\Lambda_S$ has the interpretation of a charge lattice, and the $B_i$ are associated to a 2-form gauge group $U(1)^r$, where $r$ is the number of tensor multiplets.

Let us denote a $\Z$ basis of $\Lambda_S$ by $\eta_i$, i.e. we can write any point $\eta$ in $\Lambda_S$ as
$\eta = \sum_i a_i \eta_i$ with $a_i \in \Z$. We have 
\begin{equation}
\eta_i \cdot \eta_j = \Omega_{ij} \,,
\end{equation}

where the  integer  matrix $\Omega_{ij}$ specifies intersections of the associated curves, cf. above and Appendix \ref{app:unimodularity}. 
We can use the $\eta_i$ basis to describe lattice points on the dual lattice $\Lambda_S^*$ as well:  
$\omega = \sum_i a_i \eta_i$. As opposed to $\Lambda_S$, this allows
the $a_i$ to be integer or fractional, subject to the constraint that for any $\omega \in \Lambda_S^*$, $\omega \cdot \eta \in \Z$ for all $\eta \in 
\Lambda_S$.

Wrapping $D3$ branes on $C_{\bf{a}} \times \eta$, where $C_{\bf{a}}$ is a space-time cycle and $\eta$ a (possibly non-compact) cycle in $\Lambda_S^*$ gives rise to charged surface operators in the 6D SCFT, so that $\Lambda_S^*$ also has an interpretation as a charge lattice. {Whenever $\eta$ is compact this produces a BPS string, whereas we get a defect when $\eta$ is non-compact.}

Recall that in 6D, an electric p-form  gauge symmetry has a dual $(4-p)$-form magnetic symmetry. 2-form symmetries are thus self-dual. In the type IIB setting, this is matched by the fact that $D3$ branes and the $C_4$ gauge potential are self-dual under SL$(2,\mathbb{Z}_2)$. 

The 6D tensor multiplets that contain the 2-form gauge fields $B_i$ are found, in the type IIB description, by a Kaluza--Klein reduction of $C_4$ on the non-compact base $B$. By a slight abuse of notation, we let $\eta_i$  denote 2-forms that are Poincar\'{e} dual to the above mentioned cycles in $B$, and expand  $C_4 = \sum_i B_i \wedge \eta_i$.   The coupling of the B-fields to the charged surface operators is then given by
\begin{equation}\label{eq:tensor_coupling_D3s}
\int_{D3} C_4 = \sum_{i,j} a_i \int_{C_{\bf{a}}} \Omega_{ij} B_j \, .
\end{equation}
This identifies the charges $\sum a_i \Omega_{ij}$ associated to $B_j$. Then, with $H_j = dB_j$ and $\hat{C}_{\bf{a}}$ being an  $S^3$ that links $C_{\bf{a}}$, we have that 
\begin{equation} \label{eq:diracquant}
\frac{1}{2\pi}\int_{\hat{C}_{\bf{a}}} H_i = \sum_{j} a_j \Omega_{ij}\, .  
\end{equation} 

The operators $U_g(M^3)$ for a 6D theory propagating on $M_6$ can then be written as
\begin{equation}
U_g(M_3) =  \exp\left(i \sum_j c_j \int_{M_3} H_j\right) \, ,
\end{equation}
where $c_j$ is a $U(1)$ parameter. These give the transformations of the Wilson surfaces\footnote{This generalises the standard Wilson lines, and $a_i \Omega_{ij}$ can be seen as tracing over gauge indices.  We remark that, since $a_i \Omega_{ij} b_j \in \mathbb{Z}$ for all  $b_j \in \mathbb{Z}$, these Wilson operators are invariant under large gauge transformations of the fields $B_i$, i.e. under $\int_{C_{\bf{a}}} B_j \to \int_{C_{\bf{a}}} B_j + 2 \pi b_j$. } 
\begin{equation}
 V(C_{\bf{a}}) = \exp\left(i \sum_{i,j} a_i \Omega_{ij} \int_{C_{\bf{a}}} B_j \right)\, . 
\end{equation}

This means that we can write out \eqref{eq:2fsym6Drep} as
\begin{equation}\label{eq:2fsym6Drep_explicit}
U_g(\hat{C}_{\bf{a}}) V(C_{\bf{a}}) = \exp\left(i \sum_i  c_i \int_{\hat{C}_{\bf{a}}} H_i \right) V(C_{\bf{a}}) = \exp\left(2 \pi i  \sum_{i,j} c_i \Omega_{ij} a_j\right)  
V(C_{\bf{a}}) \, .
\end{equation}
Clearly, this equation shows that $G$ is  $U(1)^r$,  where $r$ is the rank of the lattice $\Omega_{ij}$.

Now the BPS strings, which come from elements 
$\sum_i a_i \eta_i \in \Lambda_S$, where the $a_i$ are integers, break this to the defect group  \cite{DelZotto:2015isa,Bhardwaj:2020phs}.  This means that the non-broken elements of the defect group must 
act trivially in case all of the $a_i$ are integer. Thus, for any non-trivial group parameter $c_i$, we find 
\begin{equation}
 c_i \Omega_{ij} a_j  \in \Z \;\; \forall  \; a_j \in \Z \, . 
\end{equation}
Let us define $\gamma = \sum_i c_i \eta_i$. The above can be rewritten as 
\begin{equation}
\gamma \cdot \eta \in \Z  \, ,
\end{equation}
for all $\eta = \sum a_i \eta_i$. If $a_i$ is integer, so $\eta \in \Lambda_S$ we must then have $\gamma \in \Lambda_S^*$. If instead $\eta \in \Lambda_S^*$ this still gives non-trivial elements 
of the group $G_S$ of 2-form symmetries. Finally, whenever we not only have $\gamma \in \Lambda_S^*$, but the stronger condition 
$\gamma \in \Lambda_S$, the action on $\eta \in \Lambda_S^*$ is trivial on all defects coming from D3 branes wrapped on elements of $\Lambda_S^*$. Hence the 2-form symmetry group must be
\begin{equation}
 G_S = \Lambda_S^*/\Lambda_S \, .
\end{equation}

Note that the action on defects, which come from $\Lambda_S^*$, is still nontrivial, and their charges are given by the discriminant
form
\begin{equation}
\gamma \cdot \gamma' = \gamma_i \Omega_{ij} \gamma_j   \, ,
\end{equation}
for any pair $\gamma,\gamma'$ of elements in $\Lambda_S^*$, which takes values in $\Q$ mod $\Z$ \cite{MR525944}.

As we have seen above, the 2-form symmetry group is equivalent to the defect group studied in 
\cite{DelZotto:2015isa,Bhardwaj:2020phs}. As argued in these references, $G_S$ measures the screening effects of the dynamical strings in the 6D theory. The charges of defects thus constitute discrete data that must be specified to fully determine the theory.  Another way of stating the above is that specifying the 2-form symmetry group relates, through \eqref{eq:diracquant}, to distinct choices of the quantised background flux $H_j$ that take values in $H^3(M_6,\Z) \otimes \Lambda_S^*/\Lambda_S$ \cite{Tachikawa:2013hya,DelZotto:2015isa}. Now, since 3-form fluxes are self-dual in 6D, choosing such a background flux is subtle and requires a choice of duality frame. Only after such a frame is fixed, can one specify a partition function for the 6D theory \cite{Tachikawa:2013hya}, see also \cite{Aharony:1998qu,Witten:1998wy,Witten:2009at,Freed:2006yc,Henningson:2010rc,Freed:2012bs}.

\section{Coupling $(2,0)$ SCFTs to gravity}\label{sect:20andgravity}

In this section we will study how to couple 6D $(2,0)$ SCFTs to gravity. This can be accomplished by studying compactifications of IIB string theory on K3 surfaces at specific loci in the 
moduli space, see \cite{Aspinwall:1996mn} for a review.

The scalar moduli space of type IIB on a K3 surface $X$ is a Grassmanian, points of which correspond to  
positive norm five-planes $\Sigma_5$ in a vector space $\R^{5,21}$ of signature $(5,21)$, modulo the group of U-dualities:
\begin{equation}
O(\Lambda_{5,21}) \setminus O(5,21)/\left(O(5) \times O(21) 
\right)\, .
\end{equation}
We can think of $\R^{5,21}$ as containing $H^2(X,\R)= \R^{3,19}$, and the 
components of $\Sigma_5$ along this subspace as describing the integrals of 
the K\"ahler form $J_X$, (real and imaginary part of the) 
holomorphic 2-form $\Omega^{2,0}_X$, B-field $B_2$, and RR 2-form $C_2$ over 
cycles of $X$. The group of U-dualities are the automorphisms of the unique 
even unimodular lattice $\Lambda_{5,21} \cong (-E_8)^{\oplus 2} \oplus 
U^{\oplus 5}$, and it contains the automorphisms of $H^{2}(X,\Z)\cong 
(-E_8)^{\oplus 2} \oplus U^{\oplus 3}$, integral shifts of $B_2$ and $C_2$, the 
type IIB S-duality group SL$(2,\Z)$, as well as the mirror map as subgroups.

In the absence of superconformal sectors, there are furthermore $26$ 
tensor fields, out of which $21$ are self-dual and $5$ are anti self-dual. They originate 
from the KK reduction of $C_4$ along harmonic 2-forms on $X$ (this gives 22 tensors) as well as
$B_2$ and $C_2$ that give two tensors in 6D each. This 
number is uniquely fixed in 6D $(2,0)$ supergravity, as such a theory is only anomaly free when 
coupled to $21$ self dual tensors, the remaining $5$ anti self-dual tensors are 
part of the $(2,0)$ gravity multiplet \cite{Townsend:1983xt}.

Recall from the general discussion in Section~\ref{sect:6DSCFT_2form} that tensors in 6D are sourced by BPS strings, which can originate from D3-branes 
wrapped on curves in $H_{2}(X,\Z)$, the fundamental string and D1-branes in 6D,
as well as the NS5-brane and D5-brane wrapped on the whole K3 surface. As a 
consequence of U-duality, we can associate the lattice of BPS strings with the 
whole lattice $\Lambda_{5,21}$. In particular, for any such state $\eta \in 
\Lambda_{5,21}$ we can choose a U-duality frame or geometric 
interpretation\footnote{I.e. a choice of embedding of $H^{2}(X,\Z)$ into 
$\Lambda_{5,21}$ together with a choice of which directions in $\Sigma_5$ 
correspond to $J_X$, $\Omega^{2,0}_X$, $B_2$ and $C_2$.} where the state in 
questions is described by a D3-brane wrapped on an irreducible holomorphic 
curve $\eta$. Choosing a set of orthonormal vectors $V_i$ spanning $\Sigma_5$, this 
shows that its tension $T(\eta)$ (in appropriate units) is given by 
\begin{equation}
T^2(\eta) = \sum_{i=1}^5 (\eta \cdot V_i)^2 \, .
\end{equation}

The lattice $\Lambda_{5,21}$ contains various ADE root lattices $\Gamma_i$ as 
sublattices, and whenever $\Sigma_5$ is orthogonal to a root lattice 
$\Gamma_i$, the associated strings become tensionless. If there is a geometric 
interpretation where $\Gamma_i$ is contained in $H^{2}(X,\Z)\hookrightarrow 
\Lambda_{5,21}$, this implies that the associated curves\footnote{For an 
irreducible curve of class $C \in H^{2}(X,\Z)$ contained in a K3 surface, the 
self-intersection number is related to the genus by $C^2 = 2g-2$. Roots hence 
correspond to $\P^1$s.} have collapsed to zero volume forming a singularity of the corresponding ADE type, and the integrals of $B_2$ and $C_2$ over these 
curves vanish. 

The lattices of tensionless strings just discussed are the hallmarks of 6D 
SCFTs. In the 6D $(2,0)$ supergravity theory, we may hence generate 
SCFT subsectors by making sure that a root sublattice of $\Lambda_{5,21}$ is 
perpendicular to the five-plane $\Sigma_5$. As $\Sigma_5$ must be generated by 
positive norm vectors and the ADE root lattices are negative definite we can 
achieve that for any embedding
\begin{equation}\label{eq:embeddlambdaSintoLambda319}
\Lambda_S  := \bigoplus_i \Gamma_i \hookrightarrow \Lambda_{5,21} \, . 
\end{equation}
where $\Gamma_i$ is any of the ADE root lattice, $\Sigma_5$ is perpendicular to 
all of the $\Gamma_i$. Such points are furthermore finite distance in moduli 
space.
 
The question of which $(2,0)$ SCFTs can be coupled to gravity (and how) can hence be answered by 
classifying which lattices $\Lambda_S$ can be embedded into $\Lambda_{5,21}$ 
(and how). This can be achieved using the methods of \cite{Font:2020rsk}. While a complete classification is 
beyond the scope of the present work, there are a few obvious consequences that can be immediately deduced. 
For example, the sum of the ranks $r$ of superconformal sectors that can be coupled 
to gravity is at most $21$, as this is the maximal rank of a root sublattice of $\Lambda_{5,21}$ that can be orthogonal
to $\Sigma_5$. Upon circle compactification, such extremal theories have a gauge symmetry of rank $21$, which 
precisely saturates the bound of \cite{Kim:2019ths}. As detailed below, the embeddings \eqref{eq:embeddlambdaSintoLambda319} are not necessarily 
primitive, which complicates such a classification, but also leads to interesting phenomena such as gauged 2-form symmetries.

\subsection{Gauged 2-form symmetries}
\label{sec:2form20}
We are now going to take the following perspective. Assume that we have found an embedding 
\begin{equation}
\Lambda_S = \bigoplus_i \Gamma_i \hookrightarrow  \Lambda_{5,21}
\end{equation}
of a direct sum of ADE root lattices into $\Lambda_{5,21}$ and choose $\Sigma_5$ such that the 
associated BPS states give rise to tensionless strings in 6D. Decoupling 
gravity, we then get a superconformal sector that contains the associated 
$(2,0)$ SCFTs. As discussed in Section \ref{sect:2form_in_6D}, each of 
these SCFTs has a global 2-form symmetry given by the finite Abelian group
\begin{equation}\label{eq:G_S_k3}
G_S := \Lambda_S^*/\Lambda_S = \bigoplus_i \Gamma_i^*/\Gamma_i \, ,
\end{equation}
and we can think of this group as a subgroup of $U(1)^r$ which is broken to the finite group $G_S$ by the massless BPS strings. 
If we now reintroduce gravity, this introduces new BPS strings associated with the elements of $\Lambda_{5,21}$. These 
will in general transform under $G_S$ and hence break it to a subgroup $G$. Due to the absence of global symmetries 
in theories of quantum gravity, and in the absence of further extended objects that can facilitate a breaking, 
this subgroup $G$ of the group of 2-form symmetries of the SCFT subsectors must hence be gauged. 

Using the logic outlined above, we can now determine $G$. For any element $\gamma \in G$, it must be that all of the BPS 
strings associated with $\Lambda_{5,21}$ have a trivial transformation. Elements of $G_S$ are in 
one-to-one correspondence with elements of $\Lambda_S^*/\Lambda_S$, which we now think as embedded into 
$\Lambda_{5,21}\otimes \,\Q$. More 
concretely, the embedding of $\Lambda_S$ into $\Lambda_{5,21}$ and the fact that we can express $\Lambda_S^*$ using 
specific fractional linear combinations of elements of $\Lambda_S$ shows how $\Lambda_S^*$ sits inside 
$\Lambda_{5,21}\otimes \,\Q$. We then group these into orbits under $\Lambda_S$ and chose a representative.

We can express the condition that none of the BPS strings has a non-trivial transformation under $\gamma$ as 
\begin{equation}
\eta \cdot \gamma \in \Z \,\,\, \hspace{1cm} \forall \eta \in \Lambda_{5,21}\, .
\end{equation}
Hence we need $\gamma \in \Lambda_{5,21}^*$, which by the self-duality of 
$\Lambda_{5,21}$ means that 
$\gamma \in \Lambda_{5,21}$. Unbroken elements of $G_S$ are hence contained in
\begin{equation}\label{eq:G_S_K3}
G = \left( \Lambda_{5,21} \cap  \Lambda_S^*\right)/\Lambda_S  \, .
\end{equation}
As $\Lambda_S$ is embedded in the integral lattice $\Lambda_{5,21}$, it follows 
that any element of $\Lambda_{5,21}$
has integral inner form with any element in $\Lambda_S$, so that 
\begin{equation}
\Lambda_{5,21} \cap  \Lambda_S^* = \Lambda_{5,21} \cap \left( \Lambda_S \otimes 
\Q \right) \, .
\end{equation}
For a primitive embedding $\Lambda_S \hookrightarrow \Lambda_{5,21}$, $\left( 
\Lambda_S \otimes \Q \right) = \Lambda_S$, 
so that $G$ is trivial. 

Let us hence assume that the embedding $\Lambda_S \hookrightarrow \Lambda_{5,21}$ is not primitive, so that $
\tors \left( \Lambda_{5,21} / \Lambda_S \right)$ is non-trivial. We will give 
some examples of such cases 
in the sections below. For a non-primitive embedding, the torsion subgroup of 
$\Lambda_{5,21}/\Lambda_S$ is given by elements $\eta$ in $\Lambda_{5,21}$ that 
are not in $\Lambda_S$, and hence non-trivial
in the quotient, but for which a multiple $d \eta$ is in $\Lambda_S$ for $d\in \Z$ and $d>1$. This implies that $\eta$ is in 
$\Lambda_{5,21} \cap \left( \Lambda_S \otimes \Q \right)$ and hence defines a 
non-trivial element in $G$. 
Conversely, any element that is non-trivial in the quotient \eqref{eq:G_S_K3} must be in the $\Q$-span of $\Lambda_S$ without 
being in $\Lambda_S$ and hence corresponds to a torsional element in the 
quotient of $\Lambda_{5,21}$ by $\Lambda_S$. 
What we have hence shown is that the subgroup $G$ of the global 2-form symmetry group $G$ which cannot be broken by BPS strings is simply 
\begin{equation}\label{eq:this_is_G}
G = \tors \left( \Lambda_{5,21} / \Lambda_S \right)  \, .
\end{equation}
Intuitively, the presence of this torsional group means that $\Lambda_S$ is embedded in a non-minimal way in 
$\Lambda_{5,21}$, i.e. $\Lambda_S \neq \left( \Lambda_S \otimes \Q \right) \cap 
\Lambda_{5,21}$. This implies that the inner product of elements of $\Lambda_{5,21}$ with elements of 
$\Lambda_S$, which results in charges of BPS strings under the 2-form symmetries of the conformal sectors to be non-minimal as well. 

As is evident from \eqref{eq:G_S_K3}, elements of $G$ form a subset of elements of $G_S = \Lambda_S^*/\Lambda_S$. This in particular allows to determine which subgroup of $G_S$ becomes gauged upon coupling the chosen collection of SCFTs to gravity. Note that for a specific choice of $\Lambda_S$, $G$ is not unique but depends on the embedding $\Lambda_S \hookrightarrow \Lambda_{5,21}$. 

\subsection{Elliptic fibrations and non-primitive embeddings}

\label{ssec:EllipticK3} 
Having shown that the subgroup $G\subset G_S$ unbroken by BPS strings is non-trivial in cases where $\Lambda_S$ is non-primitively 
embedded in $\Lambda_{5,21}$, we now present a class of examples in which this is indeed the case. To do so, we chose a particular 
U-duality frame in which the K3 surface $X$ has an elliptic fibration. Building on the extensive literature of elliptic fibrations 
on K3 fibrations, we will show that $G$ equals the group of torsional sections when $\Lambda_S$ equals the lattice of 
fibre components not meeting the zero section. The observations of this section hence provide a fascinating application of the 
classification work of elliptic fibrations on K3 surfaces. The data specifying any of the elliptic fibrations found in the tables 
in \cite{1996293} directly translates to a possible set of $(2,0)$ SCFTs that can be coupled to 6D quantum gravity, and also directly 
tells us the gauged subgroup of the global 2-form symmetries of the superconformal sectors. Some background material concerning results 
used in this section and a guide to the original literature can be found in \cite{zbMATH05896593}.

Let us denote the K3 surface in question together with a choice of complex structure by $X$ and let us assume that $X$ has an elliptic fibration. 
Using the complex structure we can define the Picard lattice
\begin{equation}
\pic(X) = H^{1,1}(X) \cap H^{2}(X,\Z) \, .
\end{equation}
It is primitively embedded in $\Lambda_{5,21} \cong H^2(X,\Z) \oplus U^{\oplus 2}$ by construction. Together with 
the transcendental lattice $T_X = \pic(X)^\perp \in H^2(X,\Z)$, we can write 
\begin{equation}
H^2(X,\Z) \supseteq T_X \oplus \pic(X)\, .
\end{equation}
An elliptic fibration is now fixed by a primitive embedding of a copy of the hyperbolic lattice $U$ into $H^2(X,\Z)$. 
One can think about $U$ as containing the class of the fibre of the elliptic fibration and the zero section.
Given such an embedding it follows that $\pic(X)$ can be decomposed as
\begin{equation}
\pic(X) = U \oplus W \,, 
\end{equation}
where $W$ is called the frame lattice of the elliptic fibration. Denoting the fibre class by $F$, it can also be defined by $W = F^\perp/F \subset \pic(X)$. 
The frame lattice contains all of the components of reducible fibres not meeting the zero section. Such components are always $\P^1$s, 
and they span a direct sum of root lattices of ADE type. This determines a sublattice
\begin{equation}
\Lambda_S  = \bigoplus \Gamma_i \subset W \, ,
\end{equation}
The quotient 
\begin{equation}
 W/\Lambda_S = MW(X)
\end{equation}
is isomorphic to the Mordell-Weil group of the elliptic fibration. The Mordell-Weil group 
is the group of sections of the elliptic fibration, which inherits its group structure from the group law on the elliptic curve which is the fibre. 
\footnote{The identity element of this group is given by the zero section.} The Mordell-Weil theorem says that it is a finitely generated Abelian 
group, which is also clear in the present context by its identification with $W/\Lambda_S$. 
We can hence write 
\begin{equation}
MW(X) = \Z^k \oplus_i \Z_{d_i}\, , 
\end{equation}
i.e. there is a free subgroup and a torsional subgroup. 

For any elliptic K3 surface, we can go to a locus in the moduli space where all of the fibre components 
not meeting the zero section, i.e. cycles in the lattice $\Lambda_S$, are collapsed. Setting also the 
periods of the B-field and the RR 2-form $C_2$ to zero, we find $(2,0)$ superconformal sectors 
of the associated ADE types. We can now compute 
\begin{equation}
\tors \left(H^2(X,\Z)/\Lambda_S \right ) =  
\tors \left( W /\Lambda_S \right ) =  \tors\left(MW(X)\right) \, ,
\end{equation}
where we have used that $\Lambda_S$ is contained in $W$, which is primitively embedded 
in $H^2(X,\Z)$ via $\pic(X)$. 
Coupling a collection of $(2,0)$ SCFTs is hence expected to give a gauged 2-form symmetry if we embed 
the conformal sectors in a K3 surface via singular fibres, and there is non-trivial torsion in the 
Mordell-Weil group. Although from the perspective of geometry it is a non-trivial fact 
that the torsional subgroup of the Mordell-Weil group is a subgroup 
of $G_S = \oplus_i \Gamma_i^*/\Gamma_i$, our analysis starting from \eqref{eq:G_S_K3} immediately 
implies this statement as well.

It is known that an elliptic K3 surface with a fixed 
torsional Mordell-Weil group $G$ only allows for fibres with compatible monodromies e.g. 
groups with sufficiently constrained centers \cite{Hajouji:2019vxs}. These 
various K3 geometries however do not admit any arbitrary combination of possible 
$\Gamma_i$ fibres for some fixed $G$ Mordell-Weil group. E.g. for $G=\mathbb{Z}_7$ one 
finds exactly three $A_6$ but no less. This observations hints at the fact that 
there might be an 2-form anomaly appearing in this setting.

It is now easy to construct examples of this type. We can even give an algebraic model of K3 surfaces with all fibre components not meeting the 
zero section as a Weierstrass model
\begin{equation}
y^2 = x^3 + f x + g 
\end{equation}
over a base $\P^1$, and with $f$ and $g$ homogeneous polynomials of degrees $8$ and $12$. In terms of the vanishing orders of $f$, $g$ and $\Delta = 4f^3-27g^2$, 
the fibre types and lattices $\Gamma_i$ are then given in Table \ref{kodairaclass}.
\begin{table}
\begin{center}
\begin{tabular}[h]{|c|c|c|c|c|c|c|c|}
\hline
 $\ord(f)$ & $\ord(g)$ & $\ord(\Delta)$ & fibre type & $\Gamma$ &$\Gamma^*/\Gamma$  \\ \hline \hline
 $\geq 0$ & $\geq 0$ & $0$ & smooth & none & - \\ \hline
$0$&$0$&$n$&$I_n$&$A_{n-1}$  & $\mathbb{Z}_n$ \\ \hline
$2$ & $\geq 3$ & $n+6$ &  $I_n ^*$ & $D_{n+4}$  & $\mathbb{Z}_2^2$ or $\mathbb{Z}_4$ \\ \hline
$\geq 2$ & $3$ & $n+6$ &  $I_n ^*$ & $D_{n+4}$&   $\mathbb{Z}_2^2$ or $\mathbb{Z}_4$ \\ \hline
$\geq 1$ & $1$ & $2$& $II$ & none & - \\ \hline
$\geq 4$ & $5$ & $10$ & $II^*$  & $E_8$ & $\mathbb{Z}_1$ \\ \hline
$1$ & $\geq 2$ & $3$ &  $III$ & $A_1$&$ \mathbb{Z}_2$  \\ \hline
$3$ & $\geq 5$ & $9$ &  $III^*$  & $E_7$&$ \mathbb{Z}_2$ \\ \hline
$\geq 2$ & $2$ & $4$ &   $IV$  & $A_2$& $\mathbb{Z}_3$ \\ \hline
$\geq 3$ & $4$ & $8$ &  $IV^*$ & $E_6$& $ \mathbb{Z}_3$ \\ 
\hline
\end{tabular}
\end{center}
\caption{\label{kodairaclass}\textsl{The classification of reducible fibres in terms of the vanishing 
degree of $f$, $g$ and $\Delta$ and the associated ADE root lattices. The last column denotes the centres of those lattices, which for $I_n^*$ depends on whether $n$ is even or odd. }}
\end{table}
For example, for $z$ an affine coordinate on the base one may choose
\begin{equation}
 X: \hspace{.5cm} y^2 = x (x^2 - z^3 (z-1)^3 (z-i)^2 )\, ,
\end{equation}
which has two fibres of type $III^*$ corresponding to $E_7$ and one of type $I_0^*$ corresponding to $D_4$. Hence 
\begin{equation}
G_S = E_7^*/E_7  \oplus E_7^*/E_7 \oplus D_4^*/D_4 = \Z_2 \oplus \Z_2 \oplus \Z_2^{\oplus 2} \, . 
\end{equation}
Furthermore, there is a torsional section $\hat{\sigma}$ at $y=x=0$. The linear relations on $X$ imply that $2 \hat{\sigma}$ is equivalent 
to the complete intersection of $X$ with the divisor $x=0$, which translates to the fact that $\hat{\sigma}\boxplus \hat{\sigma} = 1$ in the group
law of the elliptic curve. The Mordell-Weil group hence has an element with two-torsion. As the rank of $\pic(X)$ is $20$ and hence maximal, 
the free part of the MW group must be trivial as those would contribute to the Picard group too. It follows that 
\begin{equation}
G =  \Z_2 \subset  \Z_2^{\oplus 4} = G_S \, .
\end{equation}
We furthermore have that the discriminant groups of $W$ and $T_X$ are 
\begin{equation}
G_W = G_{T_X} = \Z_2 \oplus \Z_2 \, ,
\end{equation}
which uniquely determines that $T_X$ has the inner form
\begin{equation}
T_X = \left(\begin{array}{cc}
      2 & 0 \\
      0 & 2
     \end{array}\right)\, .
\end{equation}

Working the other way, one may construct examples by starting from a choice of $T_X$ and then find the frame lattices 
and Mordell-Weil groups of possible elliptic fibrations by the method of Kneser and Nishiyama \cite{1996293}. 
A detailed review of this method can be found in \cite{Braun:2013yya} and methods for finding explicit 
Weierstrass models from such constructions (together with plenty of examples) can be found in \cite{Kumar_2014}.

\subsection{2-form symmetries in $(2,0)$ theories and fibre-base duality}
\label{ssec:20fibrebase}

In this section we want to offer an alternative perspective on the results of the last section by using a dual description 
via M-Theory. We can lift IIB string theory on a K3 surface $X$ to F-Theory by 
considering F-Theory on $Y = T^2_A \times X$ in the limit where the volume of 
$T^2_A$ goes to zero. When $T^2_A$ has finite volume, such a setup can be described by 
M-Theory compactifications on $Y$ to five dimensions. From the perspective of type IIB, this is described 
by compactification on $X \times \mathbb{S}^1$, i.e. the 6D $(2,0)$ theories considered above are reduced on a further 
$\mathbb{S}^1$.  

As in the section above, we assume that $X$ has an elliptic fibration which is such 
that its Mordell-Weil group has a non-trivial torsional component. Collapsing all fibre components not 
meeting the zero section, we can think of the resulting 6D $(2,0)$ theory as being composed of SCFT 
sectors associated with singular fibres that are coupled to gravity. As argued in the last section, 
we expect such 6D $(2,0)$ theories to have a gauged 2-form symmetry $G$. Further compactifying
this theory on a circle to a 5D $\mathcal{N}=2$ theory first yields a gauge algebra $\mathfrak{g}= \Gamma_i$
obtained due to the strings that become W-bosons when wrapping the KK circle. Furthermore, the gauged 2-form symmetry in 6D reduces 
to a gauged 1-form symmetry in 5D \cite{Bhardwaj:2020phs}. This implies that, modulo $U(1)$s,  
the 5D gauge group is $\hat{G}=\Gamma_i/G$.
 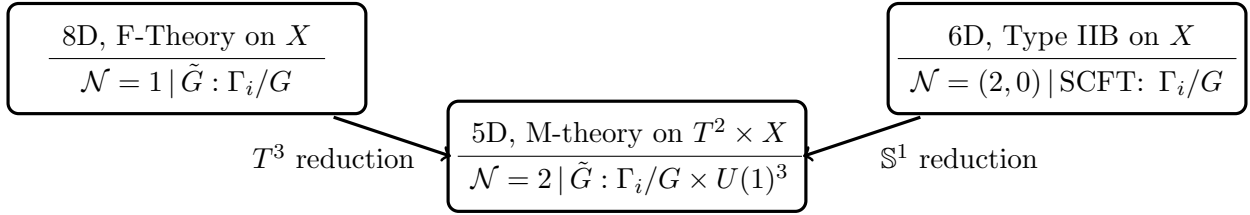
\begin{figure}[t!]
\begin{center}
	\begin{tikzpicture}[scale=1.3]
	
		\node (A) at (0,-5) [draw,rounded corners,very thick,text width=4.5cm,align=center] {  
		$ \begin{array}{c}   $5D, M-theory on $ T^2 \times X  \\ \hline
		\mathcal{N}=2\, |\, \tilde{G}: \Gamma_i / G \times U(1)^3   \end{array}$   };  
		
		\node (A) at (-4.5,-4) [draw,rounded corners,very thick,text width=4.5cm,align=center] {  
		$ \begin{array}{c}   $8D, F-Theory on $ X \\ \hline
		\mathcal{N}=1 \,|\,  \tilde{G}: \Gamma_i / G  \end{array}$   };  
	
	\node (A) at (4.5,-4) [draw,rounded corners,very thick,text width=4.5cm,align=center] {  
		$ \begin{array}{c} $ 6D, Type IIB on $ X \\ \hline
		\mathcal{N}=(2,0)\, |\,$SCFT: $\,  \Gamma_i /G    \end{array}$   };   
		
		\node (A) at (-3,-5) [ text width=4cm,align=center] {  
		$  T^3 $ reduction   };   
		
		\node (A) at (3.4,-5) [ text width=4cm,align=center] {  
		$  \mathbb{S}^1 $ reduction   };

		\draw[very thick, ->] (-3,-4.6) -- (-1.8,-5);   
		
		\draw[very thick, ->] (3,-4.6) -- (1.8,-5);   
		
		\end{tikzpicture}

\end{center}
\caption{{\it  Depiction of M-theory duality chain leading to the same 5D  
$\mathcal{N}=2$ supersymmetric theories. The same theory is obtained 
from $T^3$ compactification of F-Theory an elliptic K3, $X$ with non-simply 
connected gauge group and a circle reduction of IIB on the same singular K3 
with a gauged 2-form symmetry.   }}
\label{fig:5ddualityA}
\end{figure} 

As is apparent from the M-Theory realisation, we can also arrive at this 5D theory by first 
considering the eight dimensional theory resulting from F-Theory on the K3 surface $X$, and then 
further reducing on $T^3 = T^2_A \times \mathbb{S}^1$ to five dimensions. In the 8D F-Theory setting the non-Abelian
gauge algebra factors are again in direct correspondence to the fibre singularities 
and the global modding of the non-Abelian part of the gauge group is again induced from the Mordell-Weil 
torsion as e.g. analyzed in detail for K3 in \cite{Aspinwall:1998xj} and 
more recently also in \cite{Mayrhofer:2014opa,Hajouji:2019vxs,Cvetic:2020kuw,Apruzzi:2020zot}.
This structure persists when reducing on $T^3$, so that we again conclude that the 5D theory 
must have the same structure. From this perspective we are 
actually forced to assume that the type IIB compactification on $X$ considered initially must 
entail a non-trivial global structure, to be consistent with the 5D supergravity 
theory. Similar duality arguments can also be made for $\mathcal{N}=(1,0)$ theories 
for which multiple F-Theory lifts are possible upon circle compactification. 
We will do so in Section~\ref{sect:10fibrebase} where we will obtain a similar 
picture.

The above mentioned duality also allows us to touch upon more sophisticated 
questions that go beyond the geometric existence of the gauged 2-form 
symmetries. These include in particular a more in depth field theory analysis of
anomaly cancellation. In particular we use the 8D consistency of the center 1-form symmetry and its $T^3$ compactification as a strong evidence that also the center 2-form symmetry to which we can lift the 5D theory to be consistent. 
Gauged center 1-form symmetries in 8D supergravities have been studied from a field theory 
perspective recently in \cite{Cvetic:2020kuw}. There a new anomaly, involving the discrete 1-form symmetry
has been identified and shown that its absence restricts the embedding into the respective gauge group factors.  It is very satisfying that the very same condition has been 
found by Miranda, Person and Shimada  \cite{Miranda:1989te,2005math......5140S } in the geometry of elliptic K3's with finite MW groups. This argument shows that the 6D center 2-form symmetries of IIB on the K3's are indeed 
consistent. In more in general though this also hints at a similar 6D center 
2-form anomaly cancellation condition as in 8D to be at play which is left to be 
worked out in future work.

\subsection{Example: the mirror quartic}
  
In the last sections we have discussed how elliptic K3 surfaces with a finite Mordell-Weil group naturally lead to gauged 2-form symmetries. 
In this section, we use different techniques to give another example where 
\eqref{eq:G_S_K3} produces a finite group $G$. However we want to explicitly show two important points here: 
  First, we want to show how one can use the 
smooth geometry, i.e. the tensor branch of the SCFTs, to obtain the exact 2-form gauging and in turn to deduce
the restricted BPS string charges.   
  Secondly, we show the existence of another SCFT limit that can be taken in the same smooth geometry, that also admits a gauged 2-form symmetry by choosing a different blow-down lattice $\Lambda_S$. The second limit is chosen such, that a singular elliptic fibration with finite MW group persists unlike the first example. Therefore this example also serves as a reminder that finite MW groups are just a subset  of possibilities of how to engineer non-trivial 2-form gaugings.

 As in the section before, we work in the geometric 
setting, i.e. we chose a specific U-duality frame, a root lattice 
$\Lambda_S \subset H^2(X,\Z)$ and locus in moduli space where $\Sigma_5$ is 
perpendicular to $\Lambda_S$. As before, this means that irreducible cycles in 
$\Lambda_S$ are collapsed to zero volume and do not support non-zero $B_2$ or 
$C_2$. As $\Lambda_S$ sits purely in $H_2(X,\Z)$, and we can write 
$\Lambda_{5,21} = U^{2} \oplus H_2(X,\Z)$, and furthermore the Picard lattice is primitively embedded into $H_2(X,\Z)$, we can simplify \eqref{eq:this_is_G} to 
\begin{equation}
G = \tors(\pic(X)/\Lambda_S)\, .
\end{equation}

The example we are considering is the mirror $X^*$ of the quartic K3 surface, 
$X$ in $\P^3$. For the quartic we have that 
\begin{align}
\pic(X) = (4)  \, , \quad T_X   = (-4) \oplus U^2 \oplus (-E_8)^2 
\, ,
\end{align} 
where $T_X =  \pic(X)^\perp$ in $H^2(X,\Z) = (-E_8)^{\oplus 2}\oplus 
U^{\oplus 3 }$. For the mirror
\begin{align}
\pic(X^*) = (-4)\oplus U  \oplus (-E_8)^2  \, , \quad T_X   = (4) \oplus U
\, .
\end{align}  
This family is realized as (a resolution of) the generic 
anticanonical hypersurface in $\mathbb{P}^3/\mathbb{Z}_4 \times \Z_4$.  
The resolved ambient space is given via the fan spanned by the four homogeneous
coordinates $x_i$ of $\P^3$ and the 18 resolution divisors 
{\footnotesize
\begin{align} 
\label{eq:Mquartic}
\Delta= &\left(  \begin{array}{rrrr|rrr|rrr| }
x_0& x_1& x_2 & x_3 & f_{1,1} & f_{1,2} & f_{1,3} & f_{2,1} & f_{2,2} & f_{2,3}   \\ \hline
 -1 &-1   &-1    & 3    &-1         &-1       &-1& -1 & -1& -1   \\
 -1 &-1   & 3    &-1    &-1         &-1       &-1& 0 & 1 &2     \\
 -1 &3    & -1   &-1    & 0         &1        & 2& -1 & -1 & -1    \\ 
\end{array}\right. \nonumber
 \\ &\quad \left.    \begin{array}{|rrr|rrr|rrr|rrr|}
f_{3,1}  & f_{3,2} & f_{3,3} & f_{4,1} & f_{4,2}  & f_{4,3} & f_{5,1} & f_{5,2} & f_{5,3} & f_{6,1} & f_{6,2} &f_{6,3} \\ \hline
0   &1      &2   & -1 & -1&-1  & 0& 1    &  2  & 0 &1& 2 \\
 -1 &-1    & -1 &0    &  1&2   &-1& -1 & -1 & 2   &1 & 0\\
-1  &-1    & -1 &2    & 1 &0    &2&  1 & 0 & -1 &-1&  -1 \\
\end{array} \right)  \, .
\end{align}} 
The latter are grouped into six triples $f_{i,j}$ whose divisors resolve the respective i-th $A_3$ singularity at the vanishing of  
\begin{align}
i= 1 \ldots 6 \quad \text{ for the pairs } \quad \{ x_0 x_1, x_0 x_2, x_0 x_3,  x_1 x_2,  x_1 x_3,  x_2 x_3 \} \, .
\end{align}
The rank of the Picard lattice spanned by toric divisors as given above is 19 
dimensional, which can be computed e.g. 
via the Batyrev formula and is consistent with the expectation of mirror 
symmetry. A divisor basis is given as
\begin{align}
D_{x_3}, \qquad D_{x_1} - D_{x_3}\, ,\quad D_{x_2} - D_{x_3} \, , \qquad D_{f_{i,j}}\, .
\end{align}
As a final cross check, we compute the intersection form $\Omega$ in the above 
basis to confirm its determinant to be four, as expected from mirror symmetry.

We can use this family of surfaces to engineer a 6D $(2,0)$ theory in which 
SCFT sectors are coupled to gravity by choosing a lattice $\Lambda_S \subset 
\pic(X^*)$, and blowing down the associated curves. 

The Picard lattice $\pic(X^*)$ admits a sublattice $\Lambda_S$ of 
curves that we want to blow down parametrised by the resolution divisors 
$D_{f_{i,j}}$
\begin{align}
\pic(X^*) \supset \Lambda_S =  A_3^{\oplus 6} \, .
\end{align}
Thus the associated 2-form symmetry is given as 
\begin{align}\label{eq330}
G_S = \left(\frac{A_3^*}{A_3}\right)^{\oplus 6}= \mathbb{Z}_4^{\oplus 6 } \, .
\end{align}
In the following we claim the full possible gauged 2-form symmetry of the model 
which we obtain by collapsing the 
divisors $D_{f_{i,j}}$ inside of $\Lambda_S$ to be
\begin{align}
G= \mathbb{Z}_4 \times \mathbb{Z}_4 \subset G_S \, ,
\end{align}
and we also would like to compute their exact embedding into $G_S$. We can do so in terms of the smooth geometry, that is the tensor branch of the SCFT phase we want to consider. 
Note that the above $G$ action on the SCFT sector, is exactly the quotient action of mirror symmetry on the $\mathbb{P}^3$ ambient coordinates.
The generators that implement the above embedding can be obtained by considering 
the two linear equivalence relations
of divisors by using the dual lattice points $m_1=(0,-1,1)$ and $m_2=(1,-1,0)$ 
and formulate the linear equivalence relations
\begin{align}
\label{eq:Z4embedding}
 4 (D_{x_1} -D_{x_2}) = & (D_{f_{1,1}}+2D_{f_{1,2}}+3 D_{f_{1,3}})-(D_{f_{2,1}}+2D_{f_{2,2}}+3 D_{f_{2,3}}) + (2 D_{f_{4,1}}-2 D_{f_{4,3}}) \nonumber \\   & +(D_{f_{5,1}}+2D_{f_{5,2}}+3 D_{f_{5,3}})-(D_{f_{6,1}}+2D_{f_{6,2}}+3 D_{f_{6,3}})   \, , \\
  4 (D_{x_2} -D_{x_3}) = & -(D_{f_{2,1}}+2D_{f_{2,2}}+3 D_{f_{2,3}})+(D_{f_{3,1}}+2D_{f_{3,2}}+3 D_{f_{3,3}}) -(D_{f_{4,1}}+2D_{f_{4,2}}+3 D_{f_{4,3}}) \nonumber \\   & +(D_{f_{5,1}}+2D_{f_{5,2}}+3 D_{f_{5,3}})- (2 D_{f_{6,1}}-2D_{f_{6,3}})   \, . 
\end{align}
In the depiction above, we have grouped together the torsional relations with 
respect to the $A_3^i$-th singularities.
Here the order four modding(s) are exactly related to the fact that the prime 
divisors on the left hand side of \eqref{eq:Z4embedding} are multiplied by a 
factor four. 
   This parametrisation allows to read off the embedding of the $\mathbb{Z}_4$ quotient actions into that of the collapsed cycles in  $G_S$. 
We also give the last linear equivalence relation among the divisors relating the $D_{x_i}$ obtained from $m_3=(1,0,0)$
\begin{align}
3 D_{x_3} - D_{x_0} - D_{x_1} - D_{x_2} =  D_{f_{1,1}} & + D_{f_{1,2}} + D_{f_{1,3}} + D_{f_{2,1}} + D_{f_{2,2}} + D_{f_{2,3}} - D_{f_{3,2}}    -2 D_{f_{3,3}} \nonumber \\ 
+ D_{f_{4,1}}&+ D_{f_{4,2}} + D_{f_{4,3}}
- D_{f_{5,2}} -2 D_{f_{5,3}} - D_{f_{6,2}} - 2 D_{f_{6,3}}    \,, 
\end{align}
which together with $m_1$ and $m_2$ spans the whole dual $M$-lattice of divisor relations. As $m_1,m_2$ and $m_3$ span the whole M-lattice of the toric ambient space, these relations determine the integer second cohomology of the ambient space. One can argue using mirror symmetry that it is isomorphic to the integer second cohomology of the mirror quartic K3 surface. Computing $H^{2}(X^*,\Z))/\Lambda_S$ then amounts to setting all divisors $D_{f_{i,j}}$ to zero in the above linear relations, which yields $\Z\oplus \Z_4^2$. \footnote{One notices that the 2-form group is exactly the mirror action on the $\mathbb{P}^3$ ambient space of $X$.} 

As discussed above from the first two relations \eqref{eq:Z4embedding} one can 
read off the exact embedding 
of the $\mathbb{Z}_4$ gauging within the six $A_3$ SCFT sectors. In order to do so, we consider some BPS strings that wrap a curve $\mathcal{C}$ given by some linear combination
\begin{align}
\mathcal{C} = \sum_i a_k \mathcal{C}_k \, ,
\end{align}
with $\mathcal{C}_k$ being the curves dual to the divisors above. The BPS strings that wrap the curves $\mathcal{C}$ admit charges under the SCFT sectors given via $\mathcal{C} \cdot D_{f_{i,j}}= \lambda_{i,j} $. However note that the curve $\mathcal{C}$ must have integral intersection with
\begin{align}
\mathcal{C} \cdot (D_{x_1} -D_{x_2}) \in \mathbb{Z}  \, ,  
\end{align}
which, due to \eqref{eq:Z4embedding} restricts the $a_k$ and in turn also the BPS string charges under the $\Lambda_S$ SCFT sectors $\lambda_{i,j}$. E.g. a string charge only under the first $A_4$ can not have a weight $ \lambda_{1,1}=1$ as the only non-trivial charges but only $\lambda_{1,1}=4$. In group theory terms this tells us, that no fundamentals but four-times symmetrized representations that admits the correct center charges are allowed. All other possible representations can be deduces analogously.

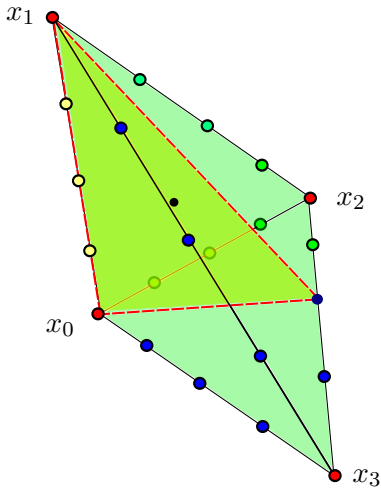
\begin{figure}[t!]
\begin{center}
\begin{picture}(0,170)
\put(-50,20){\includegraphics[scale=0.5]{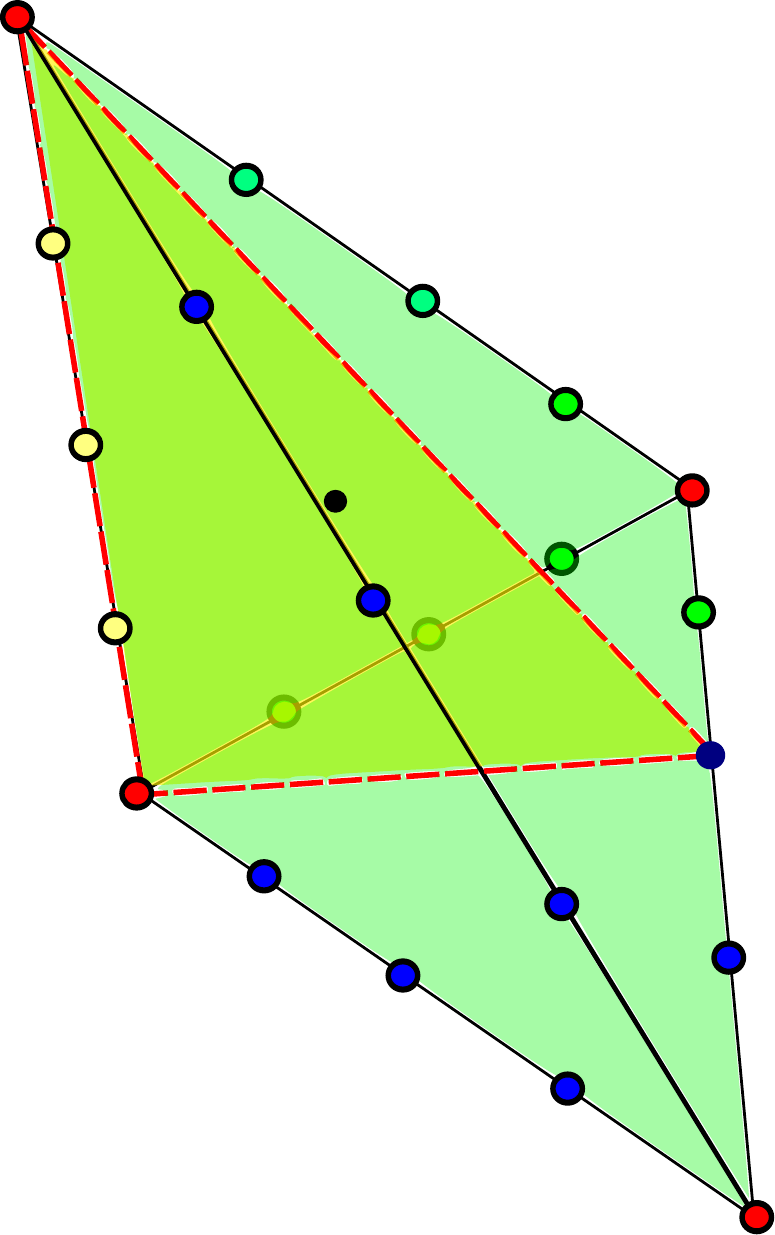}}
\put(-50,77){$x_0$}
\put(-65,195){$x_1$}
\put(66,20){$x_3$}
\put(60,125){$x_2$}
\end{picture}
\caption{{\it \label{fig:MirrorQuarticPoly}Depiction of the mirror quartic polytope. 
Edge points are given in 
red and a slice of a 2D sub-polytope, that induces an elliptic fibration 
structure is highlighted. The six $A_3$ singularities between the vertices are 
grouped with respect to the elliptic fibration. Two $\mathfrak{e}_7$ tops are 
highlighted in green and blue respectively and another $\mathfrak{su}_4$ as 
yellow points.}}
\end{center}
\end{figure}
The mirror quartic $X^*$, also admits another singular limit that gives rise to SCFTs with
gauged 2-form symmetries. These limits respect an elliptic fibration, which allows us
to connect this discussion to the general considerations made in Section~\ref{ssec:EllipticK3}.
Hence, in order to admit a non-trivial 2-form gauging, we require a non-trivial MW group.
This  structure can be seen by noting, that $X^*$ admits (six equivalent) 
sub-polytopes that are of $F_{13}$ type (in the nomenclature of \cite{Klevers:2014bqa}) that preserve a 
generic $\mathbb{Z}_2$ MW group. One choice of such an reflexive (sub-)polytope 
is given by the triangle spanned by the $x_0, x_1, f_{6,2}$ vertices. For this fibration, 
$x_0$ and $x_1$ are two sections, related by a torsional relation. One finds 
that this model admits a reducible $I_4$ fibre at a non-toric locus resolved via 
$f_{1,j}$. Furthermore one observes that the choice of the 2D sub-polytope 
slices the 3D polytope \eqref{eq:Mquartic} exactly in the middle, leaving two 
$E_7$ fibres as a {\it top} and {\it bottom} 
\cite{Candelas:1996su,Bouchard:2003bu}. 
The contributions of elliptic fibre and base, as well as all all reducible 
fibres sum up to
$2+3+2\cdot 7 =19$, which is the expected rank of the Picard lattice. The 
sublattice of those shrinkable fibres $\Lambda_S$ admits the 2-form structure
\begin{align}
G_S = \frac{SU(4)^*}{ SU(4)} \oplus \frac{E_7^*}{ E_7} \oplus \frac{E_7^*}{E_7} = \mathbb{Z}_4 \oplus \mathbb{Z}_2^{\oplus 2} \, .
\end{align}
Note that both $G_S$ and the gauged 2-form symmetry group $G$ for this choice of $\Lambda_S$ (and its embedding) are different from the choice made above, \eqref{eq330}. For this choice of $\Lambda_S$, the gauged 2-form symmetry group $G$ appears as the a $\mathbb{Z}_2$ MW torsion group of an elliptic fibration. This can be double checked by noting that the above model can be described in the most general Weierstrass model that exhibits a  $\mathbb{Z}_2$ torsion point   
\cite{Aspinwall:1998xj}
\begin{align}
y^2 = x (x^2 + a_2 x z^2 + a_4 z^4) \, ,
\end{align}
via the tuning  $a_2 =u^2 v^2\, ,  a_4 = u^3 v^3 (u+v)^2$ in the $\mathbb{P}^1$ 
base coordinates $u,v$.

\section{Coupling $(1,0)$ SCFTs to gravity}\label{sect:10andgravity} 

Having discussed $(2,0)$ theories via IIB on K3 and the gauged 2-form theories in such theories, we now adopt the same strategy for $\mathcal{N}= (1,0)$ theories using F-Theory on elliptic Calabi-Yau threefolds. Compared to $\mathcal{N}=(2,0)$ SCFTs, there are many more possibilities to construct $\mathcal{N}=(1,0)$ SCFTs, and to couple them to gravity. In particular, these constructions include various ways to enhance gauge groups, which however does not directly influence the structure of the 2-form (gauge-) symmetry that we want to focus on in this work. Therefore, we will review the structure of the geometric construction of 6D supergravity theories in F-Theory on certain bases $B$ in the following, leaving out the rich structure of tuning additional gauge groups.

\subsection{F-Theory on compact elliptic threefolds with SCFT sectors}

%%%%%%%%%%%%%%%%%%%
The vast majority of supergravity theories with eight supercharges can be described via F-Theory on an elliptic threefold $X$ over a smooth compact K\"ahler two-fold base $B$. The set of such bases $B$ is in fact not fully classified \footnote{Classification for toric and some non-toric bases can be found in \cite{Morrison:2012np,Morrison:2012js,Martini:2014iza,
Taylor:2015isa}.} and on top of this one might also tune the Weierstrass model to obtain enhanced gauge group factors over various curves. Focusing on the base, these contain in 
general curves of positive and negative self-intersection. In the type IIB description, D3 branes wrapping such curves give rise to (massive) strings in 6D with tension fixed by the volume of the curve. For $-n$ curves, which may be shrunken to yield a local geometry of the form
$\mathbb{C}^2/\Gamma$, with $\Gamma \in U(2)$, 
these strings can again become massless. Consequently, 6D supergravities admit massless string modes that are characterised by the 6D SCFT data associated to those shrunken curves. 
 
Consider a compact base $B$ and the (geometrically realized) lattice of BPS strings $\Lambda_B = H_2(B,\mathbb{Z})$. Among these we choose a specific sublattice of curves
\be 
\Lambda_S \subset \Lambda_B \, ,
\ee
that can be shrunk {\it simultaneously} creating SCFT sectors with tensionless strings. Such a choice corresponds to moving to a specific limit in the K\"ahler moduli space of the base $B$. For a single irreducible and effective curve $C$, the condition that it can be collapsed is that its self-intersection number is negative. However, as a curve is collapsed the self-intersection numbers of linked curves increase, and may become non-negative. Hence it may be difficult in general to determine possible choices of $\Lambda_S$.

Each SCFT sector potentially has a 2-form global symmetry, which must be either broken or gauged when coupling the theory to gravity, i.e. in situations with a compact base. Now, just as for $(2,0)$ theories discussed in Section \ref{sec:2form20}, if $\Lambda_S \subset \Lambda_B$ is primitively embedded, the compactification admits tensionless BPS strings that break the full 2-form symmetry $G_S$. For non-primitively embedded $\Lambda_S \subset \Lambda_B$, however, part of $G_S$ remains unbroken. Again, the unbroken elements of $G_S$  lie in
\be
G = \tors \left( \Lambda_B / \Lambda_S \right) \, ,
\ee
as can be seen by repeating the derivation given in Section \ref{sec:2form20}. Barring further symmetry-breaking objects (such as non-BPS strings), this subgroup remains as a gauged 2-form symmetry in the 6D supergravity. We will determine $G$ for different example geometries below.

Given a smooth base $B$, i.e. a $(1,0)$ supergravity theory in 6D, one may study possible blow-downs of base 
curves, find the associated lattices $\Lambda_S$, and read off which SCFT data this corresponds to. For a typical 
base $B$ there are many choices of $\Lambda_S$, each corresponding to a different instance of a collections of SCFTs coupled to gravity. Conversely, we may ask the more interesting and difficult question if a given collections of SCFTs can coexist with gravity. This is equivalent to ask if certain combinations of curves carrying 
these SCFTs can be consistently stitched together to provide a compact base manifold.

 \begin{figure}[t!]
\begin{center}
\begin{tikzpicture}[scale=1.3]
\draw[->] (0,0) -- (1,0); 
\draw[->] (0,0) -- (0,1); 
\draw[draw=red, ->] (0,0) -- (-1,1); 
\draw[draw=red,  ->] (0,0) -- (-1,-0);
\draw[draw=red,  ->] (0,0) -- (-1,-1); 
\draw[->] (0,0) -- (-1,-2); 
\draw[draw=blue, ->] (0,0) -- (0,-1); 
\draw[draw=blue, ->] (0,0) -- (1,-1); 
\node at (1.2,0) {$v_1$}; 
\node at (0,1.2) {$v_2$}; 
\node at (-1.2,1.2) {\color{red} $e_1'$};
\node at (-1.2,0) {\color{red}$e_2'$};
\node at (-1.2,-1.2) {\color{red}$e_3'$};
\node at (-1,-2.2) {$v_3$};
\node at (0,-1.2) {\color{blue} $e_1$};
\node at (1.2,-1.2) {\color{blue} $e_2$};
\end{tikzpicture} 
\end{center}
\caption{{\it A fan of a 2D toric base $B$. The fan has several 1D cones, each corresponding to a curve in $\Lambda_B = H_2(B,\mathbb{Z})$.  The red and blue 1D cones, ending in vertices $e_i'$ and $e_i$ all have negative self-intersection, and may be shrunk simultaneously to provide two SCFT sectors  $\Gamma_1, \Gamma_2$.}}
\label{fig:toric2d}
\end{figure}
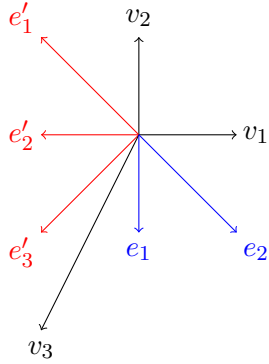

Restricting to toric base spaces offers a clear setting to address the questions raised above. When $B$ is a toric variety, the generators of the cone of curves in $\Lambda_B$ correspond to 1D rays in the fan of $B$ (see Appendix \ref{ap:cy3} for the relevant properties of toric surfaces). Blowing down a curve corresponds to deleting the associated ray while fusing the adjacent two-dimensional cones. As all cones in a fan need to be strongly convex, this singles out curves of negative self-intersections. What makes toric surfaces particular convenient is that one can immediately spot 
collections of curves that can be blown down simultaneously: such collections must be such that deleting all of the
corresponding rays respects strong convexity of the resulting fused two-dimensional cones. An example is shown in Figure \ref{fig:toric2d}. Here, $e_1$ may be shrunken simultaneously with $e_2$, but $v_1$ may not; however, we may alternatively choose to shrink both $v_1$ and $e_1$, which requires $e_2$ to stay at finite size. This selection process makes it a bit subtle to read off the possible endpoint configurations, and the corresponding SCFT sectors, from the fan of $B$.\footnote{Proceeding algorithmically one may systematically explore which end-point configurations are possible for different toric bases, and we hope to present these results in a future publication.} 
 
The toric setup we are focusing on here  allows us to reconstruct the elliptic fibration via a minimal Tate-model 
directly from the base $B$ using a simple algorithm that we discuss in Appendix~\ref{ap:cy3}.

%%%%%%%%%%%%%%%%%%%

\subsection{Extremal toric base spaces, primaries and descendants}\label{sect:prim_desc}

Blowing down curves in a toric surface will eventually result in a surface that allows no more blow-downs. This happens when we can no longer delete any rays such that the resulting adjoined cone is still strongly convex. We will call such toric surfaces extremal, which come in two types. The first type is such that the fan is composed of three rays, and the second type has four rays which are pairwise opposing. In all other cases, it follows that there are still rays which can 
be deleted by keeping all cones strongly convex. An example of an extremal base composed of three rays is given by deleting all of the rays labelled $\{e_i\}$ and  $\{e_i'\}$ from the fan shown in Figure~\ref{fig:toric2d}. 

Any toric surface is birational to $\mathbb{P}^2$ or one of the Hirzebruch surfaces $\mathbb{F}_n$  \cite{Grassi:1991ws}, i.e. we can always contract curves to reach those surfaces. Note that only $\P^2$ and $\mathbb{F}_0$ are extremal in our sense, as all other Hirzebruch surfaces can be blown down to the weighted projective space $\P_{11n}$. Although we may always blow down curves to reach these surfaces, we want to retain the freedom to choose any collapsible set of curves as $\Lambda_S$ and are particularly interested in singular bases encoding various SCFTs.

\begin{figure}[t!]
\begin{center}
\begin{tikzpicture}[scale=1.3]

\draw[->] (0,0) -- (0,1);

\draw[ ->] (0,0) -- (1,-2);

\draw[ ->] (0,0) -- (-2,-1);

\node at (1.3,-2) {$v_3$};

\node at (0,1.2) {$v_1$};

\node at (-2.2,-1) {$v_2$};

\node at (-1,0.2) {$\Gamma_1, \{\mathfrak{G}_1\}$};

\node at (1,-0.3) {$\Gamma_3,\{\mathfrak{G}_3\}$};

\node at (-0.2,-1.2) {$\Gamma_2,\{\mathfrak{G}_2\}$};

\end{tikzpicture}
\end{center}

\caption{\it A fan of an extremal base $B_o$. The fan has three rays with generators $v_i$. Each of the two-dimensional cones is furthermore labelled by a lattice $\Gamma_i$ of blown-down curves and non-minimal gauge groups $\{\mathfrak{G}_i\}$ over some curves. 
\label{fig:ext_base}}
\end{figure}
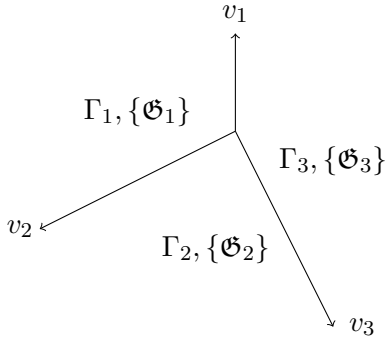

Let us hence consider blowing down a maximal set of curves $\Lambda_S$ of a smooth toric surface $B$ such that we reach a singular base $B_o$, the fan of which is composed of three rays. We will mostly discuss this case, the other extremal type can be treated analogously. We denote the ray generators by $v_1,v_2,v_3$ and the associated homogeneous coordinates by $z_1,z_2,z_3$. As any one of the blown down curves came from a ray sitting inside  of the three cones spanned by $(v_1,v_2)$, $(v_2,v_3)$ and $(v_3,v_1)$, the lattice $\Lambda_S$ can be written as the direct sum of three lattices which we denote by $\Gamma_i$
\begin{equation}
\Lambda_S  = \Gamma_1 \oplus \Gamma_2 \oplus \Gamma_3 \, .
\end{equation}
Each of these correspond to a SCFT (which may be trivial), so that we can portrait the situation by sketching the base and assigning an SCFT to each of the cones, see Figure~\ref{fig:ext_base}. The SCFT data is specified by the lattice of curves sitting in $\Lambda_S$, i.e. its charge lattice, together with possible enhancements $\{\mathfrak{G}_i\}$ of the gauge groups over some of the curves. 

Such a surface $B_o$ has three singularities $\C^2/\Z_k$ at the loci $z_1=z_2=0$, $z_2=z_3=0$ and $z_1=z_3=0$. Let us denote the triple of values $k$ for a compact toric surface by $\Xi  = [\xi_1,\xi_2,\xi_3]$. The 2-form global symmetry group of the SCFT sectors when decoupling gravity is then simply \cite{DelZotto:2015isa}
\begin{equation}\label{eq:2-formglsy3legbase}
G_S =  \Lambda_S^*/\Lambda_S = \bigoplus_i \Gamma_i^*/\Gamma_i = \bigoplus_i \Z_{\xi_i}\, .
\end{equation}

Let us start by exploiting SL$(2,\mathbb{Z})$ to put $v_1 = (1,0)$, which is always possible as ray generators must be primitive. We can then write 
\begin{equation}\label{eq:standard_fan_toric_surface}
 v_1=(1,0)\, ,\hspace{.5cm}  v_2 = (m,n)\, ,\hspace{.5cm} v_3=(l,r). 
\end{equation}
Before examining these surfaces in more detail, let us make some comments on the normal form that 
can be achieved. By distributing $v_1$, $v_2$ and $v_3$ in a counter-clockwise fashion, we have that $\det\left( v_{i} v_{i+1}\right) = |v_i| |v_{i+1}| \sin(\theta_i)$, where $\theta_i$ is the angle between 
$v_i$ and $v_{i+1}$ (we identify $v_4 =v_1$). As the cones between $(v_1,v_2)$, $(v_2,v_3)$, $(v_3,v_1)$ are strongly 
convex, the angles between these vectors must be in the range $0,\pi$, so that we find 
\begin{equation}
\det (v_1 v_2 ) = n > 0  \, \, , \hspace{1cm} \det (v_2 v_3 ) = mr -nl > 0  
\, \, , \hspace{1cm} \det (v_3 v_1 ) = -r > 0 \, ,
\end{equation}
and we can identify 
\begin{equation}
[\xi_1,\xi_2,\xi_3] = [\Delta, -r, n ] \, ,
\end{equation}
where we have set $\Delta = mr -nl$. It is important to remark that $\xi_i$ only fixes the orbifold group, but not the action on the coordinates and in turn also not the action on the fibre. This implies that different endpoints $B_o$ might have the same $\Xi$. In Section~\ref{ssec:Z7quotient} we demonstrate such an example for $\Xi=[7,7,7]$.  

It can happen that $n$ and $r$ are coprime, or that they have a common factor. If they do have a common factor $d$, $d$ must also 
divide $\Delta$. If they don't, it follows that neither of them can share a factor with $mn-nl$ as $(m,n)$ are coprime 
and $(l,r)$ are coprime. It hence follows that for any prime $d$ one of the following options must be true:
\begin{itemize}
 \item[a)] $d$ divides only one of the $\xi_i$
 \item[b)] $d$ divides all three of the $\xi_i$
\end{itemize}
i.e. it can never be true that $d$ only divides two out of the $\xi_i$.

We can use this result to describe all toric surfaces, that have a fan with $|\Sigma(1)| = 3$, as primaries and their descendants 
by global quotients. Let us first describe the primary surfaces, which are those for which $n$ and $r$ are coprime. They are fully 
described as $\C^3 - \{0\}/\C^*$ together with the weight system 
\begin{equation}
\begin{array}{ccc}
 z_1 & z_2 & z_3 \\
 \hline
 mr-nl & -r & n
\end{array}\, .
\end{equation}
As $n,r$ and $mr-nl$ are all coprime this reproduces the quotient singularities $\C^2/\Z_{\xi_i}$ by gauge fixing 
any one of the coordinates to $1$ and finding the residual finite group. 

Let us now describe the descendants for which $n$ and $r$ are not coprime. Let us write $n = d n' $ and $ r = d r'$ 
and assume that $n'$ and $r'$ are coprime. As any integer linear combination of $v_2$ and $v_3$ has the form 
$(n_1, q n_2)$ and $v_1 =(1,0)$, these ray generators do not span the 
whole N-lattice $\Z^2$, but only a sublattice $N'$ with $N/N' = \Z_d$.  By a classic result on toric morphisms \cite{Oda1988}, this implies that the toric variety $B_o$ can be written as 
\begin{equation}\label{eq:desc_global_quotient}
B_o =  \hat{B}_o/ \Z_d
\end{equation}
for a primary toric variety $\hat{B}_o$ characterised by the numbers $n'$ and $r'$. Correspondingly, we can describe $B_o$ as the quotient 
\begin{equation}\label{eq:descendant_base}
B_o = (\C^3 -\{0\}) / (\C^* \times \Z_d) \, .
\end{equation}
Note in particular that both $B$ and $B'$ have the same weight system for the $\C^*$ action.

\subsubsection*{A primary and a descendant: simple example} 
From the description \eqref{eq:descendant_base}, it is not obvious that we will indeed get singularities of type $\Z_n$ and $\Z_r$. 
Let us look at this in some more detail for an example, and consider a base $B_o$ with $\Xi = [2,4,6]$. By the above, this 
can be written as the quotient
\begin{equation}
B_o = \P_{123}/\Z_2\, . 
\end{equation}
We can realise $B_o$ from a fan with ray generators 
\begin{equation}
v_1 = (1,0)\, , \hspace{.5cm} v_2 = (-3,2)\, , \hspace{.5cm} v_3 =(3,-4) \, . 
\end{equation}
Describing this space as a quotient, we need to mod out $\C^3-\{0\}$ by (the action of) the kernel of the map 
\begin{equation}
 \phi: (t_1,t_2,t_3) \rightarrow (t_1 t_2^{-3}t_3^3,t_2^2 t_3^4) \, .
\end{equation}
Of course this contains the $\C^*$ action with weights 
\begin{equation}
\begin{array}{ccc}
 z_1 & z_2 & z_3 \\
 \hline
 6 & 4 & 2
\end{array}\, ,
\end{equation}
but we can also immediately recognise the orbifold singularities 
\begin{equation}
\begin{aligned}
 \Z_2: & (t_1,t_2,t_3) = (\zeta_2,\zeta_2,1) \\
 \Z_4: & (t_1,t_2,t_3) = (\zeta_4,1,\zeta_4) \\ 
 \Z_6: & (t_1,t_2,t_3) = (1,\zeta_6,\zeta_6) 
\end{aligned}
\end{equation}
for $\zeta_k$ a $k-$th primitive root of unity. A $\Z_2$ subgroup of the $\Z_4$ action, and a $\Z_3$ subgroup of the $\Z_6$ action 
are realized from the $\C^*$ action. They correspond to the quotient singularities present in $\P_{123}$. 

Let's describe the same situation as a $\Z_2$ quotient of $\P_{123}$. We have the weight system
\begin{equation}
\begin{array}{ccc}
 z_1 & z_2 & z_3 \\
 \hline
 3 & 2 & 1
\end{array}\, ,
\end{equation}
and the $\Z_2$ acts as 
\begin{equation}
\begin{aligned}
 &(z_1,z_2,z_3) &\rightarrow (-z_1,-z_2,z_3) \, , \\
\sim_{\lambda = \zeta_2} &(z_1,z_2,z_3) &\rightarrow (z_1,-z_2,-z_3) \, , \\
\sim_{\lambda = \zeta_4} &(z_1,z_2,z_3) &\rightarrow (\zeta_4 z_1,z_2, \zeta_4 z_3) \, .
 \end{aligned}
\end{equation}
Note that the action as described in terms of $\zeta_4$ is just $\Z_2$ due to the action of $\C^*$. The $\Z_4$ singularity is realized from a semi-direct product of two $\Z_2$'s.

\subsection{The homology lattice of an extremal surface and its resolution}

We can work out the intersection form between Weil divisors (the Chow ring) on $B_o$ by using the linear relations 
\begin{equation}\label{B_o_3_lin_rel}
\begin{aligned}
D_1 + m D_2 + l D_3 & = 0 \, , \\
n D_2 + r D_3 & = 0 \, .
\end{aligned}
\end{equation}
and the intersections 
\begin{equation}
\begin{aligned}
D_1 \cdot D_2  &= 1/n = 1/\xi_3 \, .\\ 
D_2 \cdot D_3  &= 1/\Delta  = 1/\xi_1\, ,\\ 
D_3 \cdot D_1  &= -1/r = 1/\xi_2  \, .
\end{aligned}
\end{equation}
Combining these expressions implies that 
\begin{equation}
D_i\cdot D_i = \frac{\xi_i^2}{\xi_1 \xi_2 \xi_3} \, .
\end{equation}

We can describe $H_2(B_o,\Z)$ as the quotient of $\Z^3$ by the relations \eqref{B_o_3_lin_rel}. Let us first consider 
a primary base $B_o$, i.e. all of the $\xi_i$ are coprime. The first relation in \eqref{B_o_3_lin_rel} allows us to uniquely 
express $D_1$ in terms of $D_2$ and $D_3$, so we only need to consider $\Z^2$ (spanned by $D_2$ and $D_3$) modulo $n D_2 + r D_3 = 0$. 
As $n$ and $r$ are coprime, there exist $p,q$ such that $nq-rp = 1$ and the matrix 
\begin{equation}
 M = \left(\begin{array}{cc}
            n & r \\
            p & q
           \end{array}
 \right)
\end{equation}
is in SL$(2,\Z)$ and rotates the standard $\Z$ basis of $\Z^2$ to $n D_2 + r D_3$ and  $p D_2 + q D_3$. In other words, we can write any 
lattice point in $\Z^2$ uniquely as a linear combination of these two vectors and $\Z^2$ modulo $n D_2 + r D_3 = 0$ is spanned by 
$p D_2 + q D_3$. One finds that 
\begin{equation}
( p D_2 + q D_3)^2  = \frac{1}{\xi_1 \xi_2 \xi_3} \, .
\end{equation}

For descendant basis we cannot find $p,q$ such that $nq-rp = 1$, but we can write $n = d n'$ and $r = dr'$ with $n'$ and $r'$ coprime. 
Now we can find $p$ and $q$ such that  $n'q-r'p = 1$. Hence we can write $H_2(B_o,\Z) = \Z \oplus \Z_d$, where 
the free part is generated by $p D_2 + q D_3$ and the torsion  part is generated by $n' D_2 + r' D_3$. One finds that
\begin{equation}\label{eq:square_gen_orbi}
( p D_2 + q D_3)^2  = \frac{d^2}{\xi_1 \xi_2 \xi_3} \, . 
\end{equation}

Note that these relations have precisely the right relationship that is expect from \eqref{eq:desc_global_quotient}. For a cycle dual to a class $\hat{H}$ which descends to a cycle dual to $H$ on the quotient we have 
\begin{equation}
\int_{\hat{B}_o} \hat{H}^2 = d  \int_{B_o} H^2 \, .
\end{equation}
Using our above notation we can write $\xi_i = d \hat{\xi}_i$ in this case, so that 
\begin{equation}
 \frac{1}{\hat{\xi}_1 \hat{\xi}_2 \hat{\xi}_3} =  \frac{d^3} {\xi_1 \xi_2 \xi_3} = d \frac{d^2} {\xi_1 \xi_2 \xi_3}\, ,
\end{equation}
which holds for the generators we discussed above.

We can use the above analysis to say some things about the homology lattice of a resolution $B$ of $B_o$. 
For $B$ we have that
\begin{equation}
\Lambda_B = H_2(B,\Z)
\end{equation}
is a unimodular lattice of signature $(1,h^{1,1}(B)-1)$. Let us denote the divisors which are blown down 
by $D_\mu$ and let the associated ray generators be $\nu_\mu$. The divisors $D_\mu$ span a negative 
definite sublattice $\Lambda_S \subset \Lambda_B$ of dimension $h^{1,1}-1$.

The linear relations for $B$ are
\begin{equation}\label{eq:lin_rel_resolved_base}
\begin{aligned}
 D_1 + m D_2 +l D_3  + \sum_\mu \nu_\mu^1  D_\mu & = 0 \, , \\
 n D_2 + r D_3  + \sum_\mu \nu_\mu^2 D_\mu & = 0 \, ,
\end{aligned}
\end{equation}
where $\nu_\mu = (\nu_\mu^1,\nu_\mu^2)$. 

Using the same reasoning as above, we immediately find that for primary $B_o$
\begin{equation}
  \Lambda_B / \Lambda_S = \Z \, .
\end{equation}
while for a descendant $B_o = \hat{B}_o/\Z_d$
\begin{equation}
 \Lambda_B / \Lambda_S  = \Z \oplus \Z_d\, .
\end{equation}
This implies that the embedding of $\Lambda_S$ into $\Lambda_B$ is not primitive for descendants.  
Note that it is completely irrelevant what the $\nu_\mu$ are. Whatever they are, we know that the
$\Delta_\mu$ generate all of $ \Lambda_B $ together with the $D_i$ and that the quotient 
is described by the sublattice of $\Z^3$ implied by the linear relations when the $\Delta_\mu$  
are set to zero. 

The torsion subgroup of $\Lambda_B / \Lambda_S$ is hence given by $G = \Z_d$, i.e. there is a 2-form symmetry unbroken by BPS strings which equals the common factor of the $\xi_i$. This implies that there cannot be a gauged 2-form symmetry for any primary base, while we expect such a gauging to occur for descendant bases. As descendant bases are characterised by all $\xi_i$ sharing a common factor $d$, it follows that the gauged 2-form symmetry $G$ is a diagonal subgroup of the 2-form global symmetries of the SCFT sectors \eqref{eq:2-formglsy3legbase}. 
 
\subsection{Example: unique quotient $\mathbb{P}^2/\mathbb{Z}_5$  }
\label{ssec:Z5quotient}
Here we present an explicit example of a non-crepant toric $\mathbb{Z}_5$ quotient of $\mathbb{P}^2$. This implies, that this quotient will introduce non-trivial fibres in the elliptic fibration that we will discuss. Moreover it turns out, that this quotient is unique up to an SL$(2,\mathbb{Z}_2)$ basis transformation of the toric lattice. This expectation is not a given though and will be contrasted with a non-unique $\mathbb{Z}_7$ quotient in the next section. 
 
The vertices that parametrise the toric fan of $\mathbb{P}^2/\mathbb{Z}_5$ is given as
\begin{align}
v_1 = (1,0)\, , \quad v_2= (1,5)\, , \quad  v_3=(-2,-5) \, .
\end{align}
One can easily check, that the above configuration is the unique toric quotient of that type\footnote{These vertices can be obtained from eqn.~\ref{eq:standard_fan_toric_surface} with  $n=-r=5$ and $m=-l-1=1\ldots 3$. Then there exists an SL$(2,\mathbb{Z}_2)$ transformation on the rays, that rotates the three solutions of the vertices into each other.} which supplements the $\mathbb{C}^*$ action of $\mathbb{P}^2$ by the following orbifold
\begin{align}
\label{eq:Z5patches}
\{ x_1 , x_2, x_3  \} \sim \{ \omega_5 x_1 , \omega_5^4 x_2, x_3 \} \, ,
\end{align}
with $\omega_5$ being a fifth root of unity.
   There are three orbifold singularities at the pairwise vanishing of the three associated coordinates $x_i$. The explicit local orbifold action  at each of the local patches is given as 
\begin{align}
\{ x_1 \omega_5 ,  \omega_5^{-1} x_2 \} \, , \quad \{  \omega_5 x_1, \omega_5^2 x_3   \} \, , \quad \{ \omega_5 x_2, \omega_5	^2 x_3 \} \, .
\end{align} 
Only the first patch gives rise to a crepant $A_4$ type of singularity while the other two do not, see Appendix~\ref{app:unimodularity}  for more details. Note that the $\mathbb{Z}_5$ quotient on the base can be uplifted to an action on the elliptic threefold. I.e. we start with the generic Tate model over a $\mathbb{P}^2$ base and go to a special locus in the complex structure moduli space, where the CY hypersurface becomes an invariant section under the $\mathbb{Z}_5$ action as specified by 
eqn.~\eqref{eq:Z5patches}.  This construction is further explained in Appendix~\ref{ap:cy3}. E.g. the first Tate coefficients in \eqref{eq:tatemodel} must admit the following form  
\begin{align}
a_1 =& x_3 (x_1 x_2 b_{1,1} + x_3^2 b_{1,2})   \, ,
\end{align}
with $b_{i,j}$ being some generic complex structure coefficients, to be compatible with the $\mathbb{Z}_5$ action.  

In the quotient geometry we can identify the resolution of the singular patches given in \eqref{eq:Z5patches} via equation
\eqref{eq:NonCrepant}. The resolutions are given via the linear chains 
\begin{align}
\begin{array}{c }  \\ 
 x_1\,  (2) \, (2)\, (2) \, (2)\, x_2 \end{array}\, ,  \qquad \begin{array}{c }    \mathfrak{g}_2 \, \, \mathfrak{su}_2    \\   x_1  \, (3)   (2) \,  x_3  \end{array}\, ,  \qquad\begin{array}{c}    \mathfrak{g}_2 \, \, \mathfrak{su}_2    \\   x_2  \, (3)   (2) \,  x_3  \end{array} \, ,
\end{align}
where we have supplemented intersection curves with their respective elliptic fibre singularities. Note that all three patches admit locally a global $\mathbb{Z}_5$ 2-form symmetry as expected. The resolution of the base and fibre can be 
 performed via toric geometry. It leads to a smooth threefold with Hodge numbers $(h^{1,1},h^{2,1})=(16,58)$. Using the $\frac12 ((\mathbf{7,2})+(\mathbf{1,2}))$ hypermultiplets required by the non-Higgsable clusters one can show also gauge and supergravity anomalies to be canceled.
 Finally we want to consider the gauging of the diagonal 2-form symmetry in more detail. For this we consider the explicit resolution divisors of the base, given via the following toric rays
\begin{align}
\begin{array}{llll}
f_{1,1}= (1,1) \, ,&  f_{1,2}=(1,2)  \, ,& f_{1,3}=(1,3) \, , &f_{1,4}=(1,4) \, ,   \\
f_{2,1}=(0,-1) \, ,&  f_{2,2}=(-1,-3)  \, ,& f_{3,1}=(0,1) \, ,&  f_{3,2}=(-1,-2) \, .
\end{array}
\end{align}
From those above we can derive the linear equivalence relation of the smooth base. The two relations relevant for us are given as 
 \begin{align}
 \label{eq:z5const}
  D_1   =&  2 D_3-D_2+  D_{f_{2,2}} + D_{f_{3,2}} - D_{f_{1,1}}  - D_{f_{1,2}}  - D_{f_{1,3}}  - D_{f_{1,4}} \, , \\   
  5 (D_3  -D_2)=&     (D_{f_{1,1}} +2D_{f_{1,2}}+3 D_{f_{1,3}}+4D_{f_{1,4}})-(D_{f_{2,1}}+3 D_{f_{2,2}}) +(D_{f_{3,1}} -2D_{f_{3,2}})  \, .              
 \end{align} 
 The first relation tells us, that $D_1$ can be expressed in terms of $D_2$ and $D_3$. The second relation gives the $\mathbb{Z}_5$ torsional element upon shrinking the $ D_{f_{i,j}}$ divisors. From the last relation we  also deduce the precise gauging of the $\mathbb{Z}_5$ 2-form symmetry and how it is implemented in the three SCFT sectors.
This can again be obtained by considering some curve $\mathcal{C}$ in the base that is wrapped by a BPS string. 
Lets first focus on those curves that wrap the $A_4$ part only and consider intersections with $\mathcal{C} \cdot D_{f_{1,j}}=\lambda_{j}$ with $\lambda_j$ being weights of representations of $\mathfrak{su}_5$. 
The torsion restricts those weights only to be consistent when they have a trivial $\mathbb{Z}_5$ center charge, such as the five times symmetrized fundamental representation\footnote{In terms of Young tableaux, all possible representations under the $A_4$ part are those that include zero mod 5 boxes.}.
When considering the two pairs of $(3)(2)$ clusters we find the $\mathbb{Z}_5$ gauging to act diagonal in the two factors with weights 
 \begin{align}
C \cdot \left( D_{f_{2,1}},D_{f_{2,2}} ; D_{f_{3,1}},D_{f_{3,2}}   \right) = \left( a,b ; a,b\right) \, ,  \text{ with } a,b \in \mathbb{Z}
 \end{align}
 to be allowed. This is the direct generalisation of the types of weights considered before, which however does not allow for a straightforward group theory interpretation. By using the torsion relation in \eqref{eq:z5const} however, all allowed BPS string charges can easily be deduced. 
 
\subsection{Example: non-unique quotients $\mathbb{P}^2/\mathbb{Z}_7$}
\label{ssec:Z7quotient}
As noted before, the quotient action of some $\mathbb{P}^2/\mathbb{Z}_m$ quotient is not necessarily uniquely fixed by $m$. For the order seven case there are two different configurations that we are discussing here. These are given by the vertices 
\begin{align}
v_1=(1,0) \, , \quad v_2=(p,7) \, , \quad v_3=(-p-1, 7) \, ,
\end{align}
for $p=1$ and $p=2$.\footnote{The configurations $p=3,4,5$ are related the the other two, by some  SL$(2,\mathbb{Z})$ transformation. }. Both models again admit three singular patches with SCFT sectors that can be worked out just as before. For the $p=1$ case, the resolution and their minimal fibres are given as  
\begin{align}
\begin{array}{c }  \\ 
 x_1\,  (2) \, (2)\, (2) \, (2) \, (2) \, (2) \, x_2 \end{array}\, ,  \qquad \begin{array}{c }    \mathfrak{f}_4 \, \quad \, \, \mathfrak{su}_3    \\   x_1  \, (5) (1)   (3) \,  x_3  \end{array}\, ,  \qquad\begin{array}{c}    \mathfrak{f}_4  \, \quad \, \, \mathfrak{su}_3    \\   x_2  \, (5)(1)(3) \,  x_3  \end{array} \, .
\end{align}
The full resolution of the threefold allows do compute the Hodge numbers, that are given as $(h^{1,1},h^{2,1})=(26,44)$ which matches the K\"ahler moduli and neutral singlets required for anomaly cancellation of the 6D supergravity. \\
Here we notice, that the elliptic fibration forced us to also resolve the intersection of the $(4)$ and $(2)$ curve above, to avoid a $(4,6,12)$ point at their intersection.
The second torsion type given via $p=2$ can be similarly be worked out and admits the following {\it minimal} resolutions
\begin{align} 
\begin{array}{c }    \quad \mathfrak{su}_2 \, \,  \mathfrak{g}_2    \\   x_1  \, (2)   (2)   (3) \,  x_2  \end{array} \,  , 
\begin{array}{c }    \quad \mathfrak{su}_2 \, \,  \mathfrak{g}_2     \\   x_2  \, (2)  (2)   (3) \,  x_3  \end{array} \,  ,
\begin{array}{c }    \quad \mathfrak{su}_2 \, \,  \mathfrak{g}_2     \\   x_3  \, (2)   (2)   (3) \,  x_1  \end{array} \, .
\end{align} 
The full threefold can similarly be constructed as before and admits the Hodge numbers $(h^{1,1},h^{2,1})=(20,38)$. This is again consistent with the above curve and fibre configurations as well as all anomalies.

For both configurations it is straightforward to compute the torsion relation and compute the explicit embedding into the $G_S = \mathbb{Z}_7^3$ global 2-form symmetries just as in the $\mathbb{Z}_5$ case.

\subsection{Gauged 2-form symmetries as seen from fibre-base duality}
\label{sect:10fibrebase}
In this section we want to show that it is again possible to view a gauged 2-form symmetry in terms of a gauged 1-form symmetry when compactified on a circle. This argument makes use of a fibre-base duality and is simply the extension of the argument we ran in Section~\ref{ssec:20fibrebase}  of  (2,0) to (1,0) theories. A nice consequence of this is the fact, that the reduced amount of SUSY allows for additional 6D and 5D vector and hypermultiplets that satisfy the constrained representations, expected from the higher form symmetry. 

Just as in the K3 case, this argument requires specific elliptic threefolds $X$. In particular we require $X$ to admit at least two elliptic fibrations\footnote{Threefolds with multiple torus-fibrations and their F-Theory lifts have recently been considered e.g in \cite{Anderson:2016cdu,Anderson:2017aux}.}.       
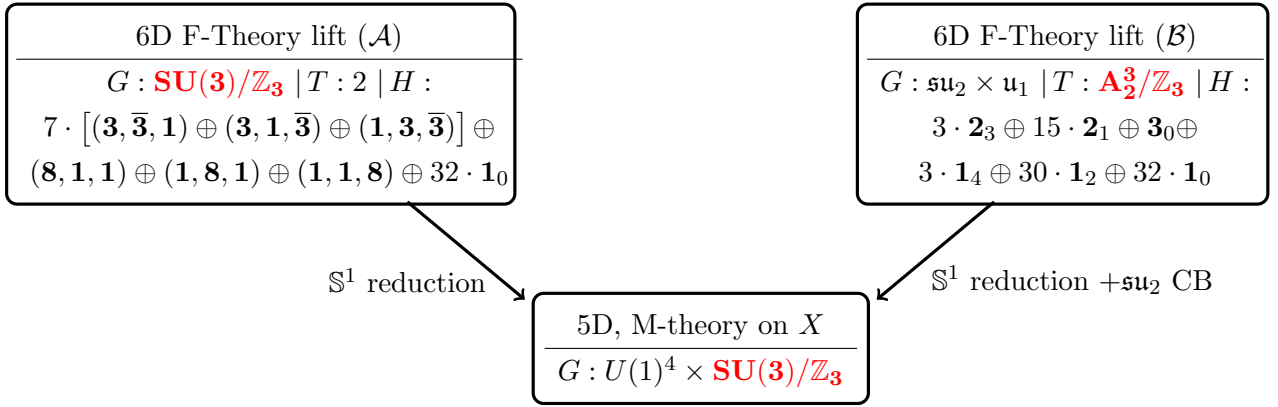
\begin{figure}[t!]
\begin{center}
	\begin{tikzpicture}[scale=1.3]
	
		\node (A) at (0,-6.5) [draw,rounded corners,very thick,text width=4.2cm,align=center] {  
		$ \begin{array}{c}  $5D, M-theory on $  X  \\ \hline
		G: U(1)^4 \times \mathbf{\textcolor{red}{SU(3)/\mathbb{Z}_3}}  \\
		      \end{array}$   };  
		
		\node (A) at (-4.5,-4) [draw,rounded corners,very thick,text width=6.5cm,align=center] {  
		$ \begin{array}{c} $6D F-Theory lift $ (\mathcal{A})  \\ \hline
		G:   \mathbf{\textcolor{red}{SU(3)/\mathbb{Z}_3}} \, \,  |\, T: 2 \, \, |\, H:  \\  
		  7 \cdot \left[(\mathbf{3,\overline{3} ,1})\oplus (\mathbf{3,1,\overline{3}})\oplus(\mathbf{1 ,3,\overline{3}})\right]\oplus \\
  (\mathbf{8,1 ,1})\oplus(\mathbf{1,8 ,1})\oplus(\mathbf{1,1 ,8}) \oplus 32 \cdot \mathbf{1}_0  
		        \end{array}$   };  
	
	\node (A) at (3.7,-4) [draw,rounded corners,very thick,text width=5.2cm,align=center] {  
		$ \begin{array}{c} $6D F-Theory lift $ (\mathcal{B}) \\ \hline
		G: \mathfrak{su}_2 \times \mathfrak{u}_1 \, \,   | \, T:  \mathbf{\textcolor{red}{A_2^3/\mathbb{Z}_3}} \,\,    | \, H: \\
		3 \cdot \mathbf{2}_{3 } \oplus 15 \cdot \mathbf{2}_{1 } \oplus \mathbf{3}_0 \oplus \\
3 \cdot \mathbf{1}_4 \oplus 30 \cdot \mathbf{1}_2 \oplus 32 \cdot \mathbf{1}_0 
		      \end{array}$   };   
		
		\node (A) at (-3,-5.8) [ text width=4cm,align=center] {  
		$  \mathbb{S}^1 $ reduction   };   
		
		\node (A) at (3.8,-5.8) [ text width=4cm,align=center] {  
		$  \mathbb{S}^1 $ reduction $+ \mathfrak{su}_2$ CB };

		\draw[very thick, ->] (-3,-5) -- (-1.8,-6);   
		
		\draw[very thick, ->] (3,-5) -- (1.8,-6);   
		
		\end{tikzpicture}

\end{center}
\caption{{\it Depiction of M-theory on $ X $ with $U(1)^4 \times SU(3)/\mathbb{Z}_3$ gauge group, that exhibits two different 6D F-Theory lifts and their massless spectra. Lift $(\mathcal{A})$ exhibits an $SU(3)/\mathbb{Z}_3$ non-simply connected gauge group in 6D. The former center symmetry becomes a gauged 2-form symmetry of the $A_2^3$  SCFT sector in the second lift 6D lift $(\mathcal{B})$.
}}
\label{fig:5dduality}
\end{figure} 
Figure~\ref{fig:5dduality} summarizes the two 6D supergravity theories and their reductions to 5D.

We start by considering an example threefold $X$ that admits the desired property, i.e. having two elliptic fibrations. The compact threefold $X$ is constructed as the anticanonical hypersurface in the ambient space $A= \mathcal{A} \times \mathcal{B}$ that is the direct product of two Fano surfaces. The two factors are given as
 \begin{align}
\mathcal{A}=  \mathbb{P}^2/\mathbb{Z}_3 \, , \qquad \mathcal{B} = \text{BL}_1 \mathbb{P}^2_{112} \, .
 \end{align}
 The Calabi-Yau hypersurface $X$ is then simply given by the divisor $[P]=c_1(\mathcal{A})+c_1(\mathcal{B})$ which is effective since both ambient pieces are Fano. However for the same reason, both ambient spaces admit an elliptic curve. 
 These tori are hence promoted to an elliptic fibre with base the other respective factor. Those fibrations of $X$ we denote by the $(\mathcal{A})$ and $(\mathcal{B})$ fibration with $\mathcal{B}$ and $\mathcal{A}$ their respective bases. 
 
Before discussing the two 6D F-Theory lifts, we consider the $\mathcal{A} = \P^2/\Z_3$ space in more detail.\footnote{This geometry has been the main example of a 6D supergravity theory coupled to SCFTs in
\cite{DelZotto:2014fia}}. In the toric description we can write the ray generators $v_i$ of the orbifold as
\begin{equation}
v_1 = (1,0)\, , \hspace{.4cm} v_2 = (1,3)\, , \hspace{.4cm} v_3 = (-2,-3) \, .
\end{equation}
We consider its resolution, which we denote by $\widehat{\mathcal{A}}$ which also amounts for a resolution of the full threefold $X$ by adding the toric vertices  
\begin{equation}
\begin{aligned}
f_{1,1} = (1,1)\, , \hspace{.3cm} & f_{1,2}  = (1,2) \, , \\
f_{2,1} = (0,1)\, , \hspace{.3cm} & f_{2,2} = (-1,-1)\, , \\
f_{3,1} = (-1,-2) \, , \hspace{.3cm} & f_{3,2} = (0,-1) \, . 
\end{aligned}
\end{equation}
These rays are associated to divisors $D_{f_{i,j}}$ which we grouped into pairs that resolve the three $\mathbb{Z}_3$ singularities each. 
 These rays admit the linear equivalence relations
\begin{equation}
\begin{aligned}
\label{eq:Z3Example}
&D_1 + D_2  - 2 D_3  + D_{f_{1,1}}  + D_{f_{1,2} } -D_{ f_{2,2}}  - D_{f_{3,1}}   = 0 \, , \\
3 &(D_3 -  D_2) =  (D_{f_{1,1}}  + 2 D_{f_{1,2}})  + (D_{f_{2,1}}  - D_{f_{2,2}})  - (2 D_{f_{3,1}}  + D_{f_{3,2}})    \, .
\end{aligned}
\end{equation}
The first relation allows to express $D_1$ in terms of $D_2$ and $D_3$ and 
 the second relation implements that $\mathbb{Z}_3$ torsional relation upon shrinking the sublattice of divisors spanned by the $ D_{f_i}$.
Hence for this piece of geometry, we admit the quotient cohomology
\begin{equation}
H^2(\widetilde{\mathcal{A}},\Z) / \Lambda_S = \Z \oplus \Z_3\, .
\end{equation} 
We can think, e.g. of $D_2$ as generating the free part, and $D_3$ as generating the torsion.  This in particular means that all holomorphic curves in the $\mathcal{A}$ part of the threefold must have constrained intersections with the divisors $D_{f_{i,j}}$. In M-theory those curves are wrapped by M2 branes leading to massless and massive particles in 5D. For this we repeat the usual argument as before: A  curve $\mathcal{C}$ in $X$ must have restricted charges under the collapsing divisors $D_{f_{i,j}}$ in order to fulfil the torsion relation. I.e. we have in general the weights \begin{align}
\mathcal{C}\cdot D_{f_{i,j}} =   \lambda_{i,j} \, ,
\end{align}
 under the $i-$th $A_2$ type of singularity. As one can easily read off, only those weights such as $ \lambda_{i,j}= (3m,0;0,0; 0,0)$ or $\lambda=(1,0;-1,0;0,0)$ are allowed where the $;$ splits up the $i-$th $A_2$ block. In terms of SU(3) representations, these are $3n$ symmetric representations or bi-fundamental representations respectively that are allowed. Hence the $\mathbb{Z}_3$ factor acts diagonally in the three $A_2$ factors. This observation can also be made more concrete, when considering the two 6D lifts in more detail.

We start by discussing the fibration $(\mathcal{A})$ that is the 6D F-Theory lift of the fibre that is embedded as the anticanonical hypersurface in $\mathcal{A}$ with $\mathcal{B}$ being its base. This fibration can be viewed as a restricted cubic where the three $A_2$ ambient singularities become $I_3$ fibres. This fibration admits a section $D_2$ but also another torsional section at $D_3$. In fact the second relation in \eqref{eq:Z3Example} maps to the fibral part of the torsion Shioda map\cite{Mayrhofer:2014opa,Oehlmann:2016wsb} of the elliptic fibration. The global 6D gauge group is therefore of the form $SU(3)^3/\mathbb{Z}_3$. A deeper analysis of the spectrum is given in \cite{Klevers:2014bqa} which confirms the above M-theory expectation. Here one finds that some of the M2 branes lift to massless hypermultiplets in 6D states that arise from codimension two singularities of the elliptic fibre. More important, we do not find massless fundamentals under any single $SU(3)$ factor.
The spectra of both 6D F-Theory lifts are given in Figure~\ref{fig:5dduality} which can be used to confirm cancellation of all 6D supergravity and gauge anomalies. 

We now discuss the $(\mathcal{B})$ fibration which admits $\mathcal{A}=\mathbb{P}^2/\mathbb{Z}_3$ as its base. We wont focus much on the fibral part, i.e. the $\mathfrak{su}_2 \times \mathfrak{u}_1$ gauge symmetry and its matter as it is not of much relevance in the following. An important side remark though is, that the $\mathfrak{su}_2$ divisor is a genus-one curve in the base in the class $c_1(\mathcal{A})$ which hence, does not intersect the three $A_2$ factors\footnote{Similarly the $\mathfrak{u}_1$ divisor does not intersect any of the singular loci either.}. Such a non-trivial intersection could have led to a non-trivial gauging of the SCFT sectors which is not present here.
 
The main point of this models is, that there are three $A_2$ SCFT sectors in the base that admit an overall diagonal gauging enforced via the torsional $\mathbb{Z}_3$ action implemented by \eqref{eq:Z3Example}. Analogously to the $(\mathcal{A})$ fibration, it acts diagonally in the global $G_S=\mathbb{Z}_3^3$ 2-form symmetry group of the three$A_2$ sectors.  This conclusion is enforced when compactifying to 5D and going to the $\mathfrak{su}_2 \times \mathfrak{u}_1$ Coulomb branch in order to match both theories. Wrapping the massless BPS strings on the circle then results in an $SU(3)^3/\mathbb{Z}_3$ gauge symmetry in 5D, matching it to the circle compactification of the $(\mathcal{B})$ theory. 

Similar as for the (2,0) fibre-base duality, we expect to have a consistent chain of theories and their dimensional reductions. I.e. from the consistent 6D $SU(3)^3/\mathbb{Z}_3$ gauge theory we expect a consistent 5D circle reduced theory with the same gauge group and in turn a well behaved 6D uplift to theory $(\mathcal{B})$ with a gauged 2-form symmetry. However, it would be very interesting to analyze those 2-form symmetries in 6D and 5D from the field theory perspective as e.g. done in \cite{Heidenreich:2021tna,Cvetic:2021sxm}. 

\subsection{Little strings with gauged 2-form symmetries}
Finally, we want to show that we can engineer also gauged 2-form symmetries within LSTs i.e. non-gravitational theories 
in the same spirit as before. Similar as for the gravitational theories we simply need is a compact (sub-)lattice $\Lambda_B$ into which the shrinkable curves $\Lambda_S$ are non-minimally embedded. As a starting point we consider an elliptic fibration with a simple rank one LST base $B_l=\mathbb{P}^1 \times \mathbb{C}$. In analogy to the compact theories we expect the LSTs from bases $\hat{B}_l=B_l/\mathbb{Z}_m$ to admit a gauged $\mathbb{Z}_m$ 2-form symmetry. 

The argument goes similar as in the compact examples. For this we describe the non-compact base by toric geometry via the fan  generated by the three vertices
\begin{align}
v_1 = (1,0) \, , \qquad v_2 =(-1,0)\, , \qquad v_3=(p,m) \, ,
\end{align}
with $p,m$ being co-prime\footnote{Note that this geometry can be seen as the decompactification of a supergravity theory with the same gauged 2-form symmetry of type $\mathbb{F}_0/\mathbb{Z}_m$ by simply adding the ray $v_4=(-p,-m)$ to the configuration.}. In the above toric diagram, there are two singular patches with in general two non-crepant $\mathbb{Z}_m$ singularities. Those are resolved as usual, by adding the vertices $v_\mu=(a_\mu,b_\mu)$ to split the two singular cones.  
This leads to the two following linear equivalence relations, 
\begin{align}
D_1  - D_2 + &p D_3 + \sum_\mu a_\mu D_\mu   = 0 \, , \\
m &D_3 + \sum b_\mu D_\mu = 0 \, ,
\end{align}
with $D_\mu$ being the divisors that can be collapsed back to the singular geometry. From the above relation one first finds $D_1$ to be linear equivalent to the generator $D_2$ and $D_3$ to be pure torsion when collapsing the $D_\mu$. The cohomology of the smooth base modulo the resolution divisors yields
\begin{align}
H^2 (  (\hat{B}_l)_{\text{res}} , \mathbb{Z})/\langle D_\mu \rangle = \mathbb{Z} \oplus \mathbb{Z}_m \, .
\end{align}
Hence the gauged 2-form symmetry is again just the torsion part in the above quotient cohomology and hence $G=\mathbb{Z}_m$. The simplest class of example for such little strings can be engineered by taking  $m=2\ldots 12$ and $p=1$. The minimal resolution is the given via the following linear chain 
 \begin{align}
 (m) (1) \underbrace{(2) \ldots (2) }_{ \times (m-1)}\, ,
 \end{align}
 that is a non-Higgsable cluster glued to an $A_{m-1}$ theory via an E-string. 
 By computing the Smith normal form of the above type of theories, it is straightforward to show triviality of the {\it global} 2-form symmetry \cite{Bhardwaj:2020phs}.  
 Note that the above little string configurations really just include the minimal resolutions of the base. Enhancing the geometry to an elliptic fibration, we might also need to blow-up at some generic points in order to avoid $(4,6,12)$ points in the Weierstrass model.  In particular for $m=9 \ldots 11$ this requires to attach $3\ldots 1$ E-string curve(s) to the $(m)$ curve. Those additional resolutions however, do not obstruct our argument when the are collapsed to the quotient base. 

\section{Discussion and outlook}\label{sect:outro}

In this work we have studied how to couple 6D $(2,0)$ and $(1,0)$ SCFTs to gravity. In both cases, the question of which SCFTs can coexist with quantum gravity can be addressed by identifying an appropriate sublattice $\Lambda_S$ of the charge lattice $\Lambda_B$ of the 
theory. The lattice $\Lambda_S$ is the identified with the lattice of BPS strings of the SCFT sector. 

For $(2,0)$ theories, $\Lambda_B$ is uniquely given by the even unimodular lattice $\Lambda_{5,21}$, and $\Lambda_S$ is a direct sum of ADE root lattices. Any such $\Lambda_S$ that allows an embedding into $\Lambda_{5,21}$ then corresponds to an instance of coupling the associated SCFTs to gravity. This allows for a full classification by lattice theoretic techniques. 

For $(1,0)$ theories, the situation is more complicated in several ways. First of all, $\Lambda_B$ is now identified with $H^2(B,\Z)$ for $B$ the base of an elliptically fibered Calabi-Yau threefold, for which are many different choices and no complete classification exists. Also, the lattice $\Lambda_S$ is now not only composed of ADE root lattices but can have other summands appearing in the classification of \cite{Heckman:2013pva,Heckman:2015bfa}. Finally, just specifying $\Lambda_S$ as a sublattice of $\Lambda_B$ is in general not sufficient, as one needs to make sure that it is generated by effective and irreducible curves that can be simultaneously collapsed. We addressed these issues by restricting our study of $(1,0)$ theories to the case of toric bases. For each such base, it is straightforward to identify which lattices of effective curves $\Lambda_S$ can be collapsed. In particular, we focused on the endpoints of such blow-downs for which no further curves can be collapsed.

The central result of this work was to examine the fate of the 2-form global symmetries $G_S$ of the SCFT sectors. We showed that the subgroup $G$ of these 2-form symmetries that remains unbroken by BPS strings is non-zero precisely when $\Lambda_S$ is non-primitively embedded into $\Lambda_B$, in which case $G  = \tors(\Lambda_B/\Lambda_S)$. 

That $G$ becomes a gauged 2-form symmetry in this case can also be argued by duality for many examples. The crucial idea is to go to five dimensions where the gauged 2-form symmetry implies a gauged 1-form symmetry. For appropriate choices of $\Lambda_S$ and for cases where the geometry in question has a second elliptic fibration, we then recovered the gauged 1-form symmetry from well-known results about torsional Mordell-Weil groups \cite{Aspinwall:1998xj,Mayrhofer:2014opa}. 

Given the simplicity of the derivation of our result, it seems natural to conjecture that it can be generalised to similar situations. Whenever charge lattices of subsectors with p-form symmetries are non-primitively embedded into the charge lattice of the full supergravity theory, there should be an unbroken subgroup that becomes gauged. A simple such case is Narain compactification of the heterotic string. It is a classic result that Narain compactification \cite{NARAIN198641,NARAIN1987369} gives enhanced gauge symmetries whenever there are roots in the even self-dual lattice $\Lambda_{d,16+d}$ that are perpendicular to the $d$-plane $\Sigma_d$ in $\R^{d,16+d}$ that fixes the location in moduli space. The lattice $\Lambda_S$ generated by such roots is then a direct sum of lattices of ADE type, and determine the algebra of the gauge enhancement. However, the embedding of $\Lambda_S$ into $\Lambda_{d,16+d}$ does not need to be primitive, i.e. the orthogonal complement of $\Sigma_d$ in $\Lambda_{d,16+d}$ can have generators that are not roots. In such cases we hence expect a gauged 1-form symmetry, i.e. the resulting gauge group of such models is the quotient of a product of simply connected ADE Lie groups by a subgroup of the center that is isomorphic to $\tors(\Lambda_{d,16+d}/\Lambda_S)$. \footnote{In fact, this is also a feature of the heterotic $SO(32)/\Z_2$ string, for which the root lattice $\Lambda_S = D_{16}$ is non-primitively embedded into the even and self-dual lattice $D_{16};\Z_2$, the charge lattice of the $SO(32)$ heterotic string in 10D. Crucially, $D_{16};\Z_2$ results from $D_{16}$ by adding in a `glue vector' which is not a root \cite{conway1998sphere}.} Building on a recent general discussion of how to classify such embeddings \cite{Font:2020rsk}, the global structure of gauge groups (i.e. the gauged 1-form symmetries) in Narain compactification were indeed found to follow this logic in \cite{Font:2021uyw,Fraiman:2021soq}. In the case $d=5$, this also follows from our discussion in Section \ref{sect:20andgravity} upon reduction on $\mathbb{S}^1$.

Having worked out the geometric origin of such gaugings, there are several directions to extend this work. In particular it would be interesting to consider the field theory consistency conditions of those symmetries, analogous to 1-form symmetries \cite{Cvetic:2020kuw,Apruzzi:2020zot} for $(1,0)$ and $(2,0)$ theories. Indeed, when discussing such symmetries we have always obtained multiple SCFT sectors upon which the  gauging acts diagonally, similar as in 1-form symmetries in 8D \cite{Cvetic:2020kuw}. Another hint for such a mechanism at play, as discussed in Section~\ref{ssec:20fibrebase}, is the fact that both such theories reduce to the same 5D theories upon circle reductions and hence have to fulfil related anomaly cancellation conditions.  

Closely related to anomaly cancellation is the question of what the maximal orders of the gauged 2-form symmetries might be, and how many independent factors can be present. In the case of $(1,0)$ such a classification required to know all F-Theory bases, which is not known except for toric and some non-toric cases \cite{Morrison:2012js,Martini:2014iza,Taylor:2015isa}. For the toric case, this question can be answered using the methods presented in this work, which we will report on in the near future. For $(2,0)$ theories on the other hand, this classification can be obtained by studying embeddings of direct sums of root lattices into the $\Lambda_{5,21}$ using the algorithm outlined in \cite{Font:2020rsk}. The classification of which combinations of $(2,0)$ theories can be coupled to gravity and which gauged 2-form symmetries result from this also allows to classify 4D $\mathcal{N}=4$ supergravity theories with non-simply connected gauge groups and their possible theta angles, extending work of \cite{Kim:2019ths}.  

Another interesting direction is to analyze the imprint of gauged 1-form 
and 2-form symmetries on enumerative invariants of the geometry. E.g.
in the dual M-theory matter states are obtained from M2 branes\footnote{Such techniques are closely 
related to F-Theory as e.g. has been used in 
\cite{Paul-KonstantinOehlmann:2019jgr,Kashani-Poor:2019jyo,Grassi:2021wii}. } that wrap
curves in the geometry, counted by Gopakumar-Vafa (GV) \cite{Gopakumar:1998ii,Gopakumar:1998jq} invariants. Since these states are restricted by the higher-form symmetry, so must be the GV 
invariants of the elliptic threefolds. Furthermore, it has been shown for gauged 
1-form symmetries on a circle \cite{Grimm:2015zea} that the theory inherits a symmetry under 
certain fractional large gauge transformation. We expect a similar symmetry to hold among the 
5D vectors that are obtained from the 6D tensor multiplets.

\section*{Acknowledgements}
We thank  Luca Cassia, Markus Dierigl and Thorsten Schimannek for interesting discussions as well as 
Robin Schneider and Matthew Magill for initial collaborations. 
The research of M.L. is financed by Vetenskapsradet under grant number 2020-03230.
The work of P.K.O.  is supported by a grant of the Carl Trygger Foundation for Scientific Research. 
This project was initiated with support from the Uppsala-Durham joint Seedcorn Fund.  
\begin{appendix}

\section{Lattices}\label{app:unimodularity} 
 
In this appendix we collect a few properties of lattices and lattice embeddings that are used throughout 
the text. For a more detailed review we recommend \cite{MR525944}. 

\subsection*{Lattices and discriminant forms}

We will use the term \emph{lattice} $\Lambda$ to refer to a finitely generated free Abelian group together with an integral bilinear 
form $\cdot$, i.e. for all $l,l' \in \Lambda$, $l\cdot l' \in \Z$. Here, the word \emph{free} means that 
$n l \neq0$ for every $l \neq 0$ and all $n \in \Z$ with $n \neq 0$. This implies that as an Abelian group (i.e. forgetting 
the bilinear form) $\Lambda \cong \Z^r$. Choosing a $\Z$-basis $\{l_i\}$ of $\Lambda$ we can write the bilinear 
form as $l_i \cdot l_j = \Omega_{ij}$. If the rank of the matrix $\Omega$ is $r$, the difference between 
positive and negative eigenvalues of $\Omega$ is called the \emph{signature} of $\Lambda$. A lattice is called 
\emph{even} if $l \cdot l \in 2\Z$ for all $l \in \Lambda$ and \emph{odd} otherwise.

By tensoring with the reals $\Lambda_\Q := \Lambda \otimes \Q$ becomes a 
vector space, and the bilinear form between lattice elements naturally extends to $\Lambda_\Q$ 
The \emph{dual lattice} $\Lambda^*$ is the subset of $\Lambda_\Q$ that has an integral product 
with all elements of
$\Lambda$:
\begin{equation}
\Lambda^* = \{ \ell \in \Lambda_\Q  | \ell \cdot l \in \Z \,\,\forall l \in \Lambda\} \, .
\end{equation}
We can use the basis $\{l_i\}$ to express elements of $\Lambda^*$ as well, but then the coefficients will in general 
not be integer, but rational numbers. As $l\cdot l' \in \Z$ for all $l,l' \in \Lambda$, it follows that $\Lambda \subseteq \Lambda^*$. When $\Lambda^* = \Lambda$
the lattice $\Lambda$ is called self-dual or unimodular. This implies that $\det(\Omega) = \pm 1$. Such matrices 
are called unimodular as well. A simple example of an even unimodular lattice is given by the \emph{hyperbolic lattice} $U$ with inner form
\begin{equation}
U = \left(\begin{array}{cc}
     0 & 1 \\
     1 & 0
    \end{array}\right) \, .
\end{equation}
This is the unique (up to isomorphism) even unimodular lattice of signature $(1,1)$. 

As $\Lambda \subseteq \Lambda^*$ we can consider the quotient 
\begin{equation}
G_{\Lambda} :=  \Lambda^*/\Lambda \, ,
\end{equation}
which is called the \emph{discriminant group} of $\Lambda$. As $\Lambda^*$ is contained in  $\Lambda_\Q$, we can 
extend the bilinear form to $\Lambda^*$ (where it ceases to be integral in general) and hence to $G_{\Lambda}$. 
For $\gamma,\gamma' \in G_{\Lambda}$ we have that 
\begin{equation}
q_\Lambda(\gamma,\gamma') = \gamma \cdot \gamma'\,\ \mbox{mod}\,\, \Z  \, ,
\end{equation}
which is called the \emph{discriminant form} of $\Lambda$.  

\subsection*{ADE root lattices}

As an important class of examples, consider the ADE root lattices $\{ A_n, D_n, E_6,E_7,E_8\}$. 
We shall use the conventions natural in geometry, where they are 
negative-definite. As each of these is generated by simple roots, which square to $-2$, these are all even lattices.
The dual lattices are the weight lattices and the discriminant groups and forms are 
\begin{equation}
\begin{array}{c|c|c}
\Gamma & G_{\Gamma} & q_{\,\Gamma} \\
\hline
A_n & \Z_{n+1} & - n/(n+1) \\
D_{2n} & \Z_2 \times \Z_2 & \left(\begin{array}{cc}
                           -n/2 & -(n-1)/2 \\
                            -(n-1)/2 & -n/2
                          \end{array}\right)\\
D_{2n+1} & \Z_4 & - (2n+1)/4 \\
E_6 & \Z_3 & -4/3\\
E_7 & \Z_2 & -3/2 \\
E_8 & - & -
\end{array}
\end{equation}

\subsection*{Embeddings and orthogonal complement}

For a sublattice $\Lambda \subset M$ the embedding of $\Lambda$ is called \emph{primitive} if the quotient
$M/\Lambda$ is free, i.e. is again a lattice. This implies that for every $\ell \in M$ such that 
$\ell \notin \Lambda$, it cannot happen that there is an $n \in \Z$, $ n\neq 0$, such that $n \ell \in \Lambda$, as 
this would imply that $\ell \neq 0$, but  $n\ell = 0$ in the quotient. Primitivity of an embedding is equivalent to 
$M \cap \Lambda \otimes \Q = \Lambda$. For non-primitive embeddings, the quotient $M/\Lambda$ contains finite groups, which 
are called the torsional subgroup $\tors(M/\Lambda)$.

For any embedding, we may consider the \emph{orthogonal complement} 
\begin{equation}
\Lambda^\perp = \{\ell \in M | \ell \cdot l = 0 \,\, \forall \, l \in \Lambda\} \,. 
\end{equation}
The orthogonal complement is automatically primitively embedded in $M$.

In case both $\Lambda$ and $\Lambda^\perp$ are primitively embedded into an even unimodular lattice $M$, 
it follows that 
\begin{equation}
q(\Lambda) = - q(\Lambda^\perp) \, ,
\end{equation}
so that in particular $G_\Lambda \cong G_{\Lambda^\perp}$. 

Note that even if $\Lambda$ is primitively embedded in some (not necessarily unimodular) lattice 
$M$, the relation 
\begin{equation}
M \supseteq \Lambda \oplus \Lambda^\perp  
\end{equation}
does not necessarily become an equality. The exception is the case when $\Lambda$ is a unimodular lattice, in which 
case the above becomes an equality.

\subsection*{Poincar\'e duality and unimodularity}
 
In this section we review the well-known fact that Poincar\'e duality for a complex surface $S$
implies that the inner form between $2$-forms is a unimodular lattice. The proof is essentially the same
as the one showing that self-duality implies unimodularity. 

The integral cohomology $H^2(S,\Z)$ is general not a free Abelian group, as it can contain torsion. Let 
us denote the free part by $H^2_f(S,\Z) := H^2(S,\Z)/\,\mbox{tors} \, \left(H^2(S,\Z))\right)$.  
Poincar\'e duality can then be stated as the map 
\begin{equation}
H^2_f(S,\Z) \rightarrow \mbox{Hom}(H^2_f(S,\Z),\mathbb{Z}) 
\end{equation}
being an isomorphism. Let us choose a $\Z$-basis $\{l_i\}$ of $H^2_f(S,\Z)$. We can then write any  
element of $H^2_f(S,\Z)$ as $\gamma = \sum_k a_k l_k$ for $a_k \in \Z$. We can choose a basis $\ell_j$ of 
$\mbox{Hom}(H^2_f(S,\Z),\mathbb{Z})$ by
\begin{equation}
v_j: \gamma \rightarrow a_j \, .
\end{equation}
Poincar\'e duality now implies that we can identify these with a basis $\{\ell_j\}$ of 
$H^2_f(S,\Z)$ which must satisfy 
\begin{equation}
l_i \cdot \ell_j = \delta_{ij} \, .
\end{equation}
As $l_i$ is a $\Z$-basis of $H^2_f(S,\Z)$ we can write
\begin{equation}
\ell_j = \sum_i B_{ji} l_i 
\end{equation}
for integers $B_{ji}$.

We can now work out that $\Omega_{ij} = l_i \cdot l_j$ is a matrix of determinant $\pm 1$. Consider
\begin{equation}
\delta_{ij} = l_i \cdot \ell_j = l_i \cdot \sum_k B_{jk} l_k = \sum_k \Omega_{ik} B_{jk}   \,.
\end{equation}
As the matrix on the left hand side has determinant $1$, and $\Omega$ and $B$ are matrices with integer entries, 
they must both have determinant $\pm 1$. Hence the inner form $\Omega_{ij}$ is a unimodular matrix. 

\subsection*{Smith normal form and discriminant group}

For any matrix $\Omega$ with integer entries, one may construct its \emph{Smith normal form} \cite{swinnerton-dyer_2001,stein_ant}. 
The Smith normal form of an integer matrix $M$ is the unique diagonal integer matrix 
$D = \diag(\alpha_i)=N\Omega S$ for invertible integer matrices $N$ and $S$, with 
increasing numbers $\alpha_i$ such that $\alpha_i \mid \alpha_{i+1}$. The matrices $N$ and $S$ are a composition of
elementary row and column operations, and as such have determinants $\pm 1$.

We first show that the Smith normal form of a matrix $\Omega$ is trivial if and only if 
$\Omega$ is unimodular. As we have shown the existence of a matrix $B$ with integer 
entries such that 
\begin{equation}
\id = \Omega \, B \, ,
\end{equation}
for this case, and $\id$ has all of the properties of the Smith normal form, it follows that the Smith normal 
form of a unimodular matrix is the identity matrix $\id$. The converse of this also holds. Let 
$\id = NAS$ for integer matrices $N,S$ and $A$. Taking determinants of both sides, we need all three
matrices $N,S,A$ to have determinant $\pm 1$, so that they are all unimodular. Alternatively, one may use 
that $N$ and $S$ are unimodular to see that the Smith normal form of a unimodular matrix is $\id$.

Let us now consider the case of lattices that are not self-dual. Using the structure theorem of finite 
Abelian groups, we can write
\begin{equation}
\Lambda^*/\Lambda = \bigoplus \Z_{p_i^{n_i}} = \bigoplus_j \Z_{k_i} \, .
\end{equation}
Here, the $p_i$ are primes (a prime can appear more than once) and the $n_i$ are integers. The alternative
presentation on the rhs uses integers $\alpha_i$ such that $k_i \mid k_{i+1}$. Furthermore, these numbers 
are unique. Let us use vectors $l$ with arbitrary integer components $l_i$ to describe $\Lambda$. 
The dual lattice $\Lambda^*$ is then composed of all $\ell$ such that 
\begin{equation}
\Omega \ell  \in \Z^r \, .  
\end{equation}
As the Smith normal form of a lattice $\Lambda$ with bilinear form $\Omega$ is $N\Omega S$ with $N$ 
and $S$ elementary row and column operations, we may equivalently describe $\Lambda^*$ as the set of 
vectors $\ell$ such that 
\begin{equation}
\diag(\alpha_i) \ell \, \in\, \Z \, .
\end{equation}
Hence, $\Lambda^*$ is generated by $\ell_i = \alpha_i^{-1}$. As $\Lambda$ is generated by $l_i =1$, 
this implies that 
\begin{equation}
 \Lambda^*/\Lambda = \bigoplus_j \Z_{\alpha_i}\, ,
\end{equation}
and we see that the Smith normal form contains equivalent information to the discriminant group. 

\subsection*{Toric singularities and continued fractions}

For toric surfaces, the types of quotient singularities are $\C^2/\Z_p$. We can describe the different actions of $\Z_p$ on $\C^2$
by working out which fans give rise to such singularities. In toric geometry, $\C^2/\Z_p$ is described by a two-dimensional fan that has
two rays and one two-dimensional cone. Denoting the ray generators by $v_1$ and $v_2$, we can use SL$(2,\mathbb{Z})$ to set $v_1 = (1,0)$.
Then, $v_2$ must have the form $v_2 = (q,p)$ with $q$ and $p$ coprime. The group action is then given by the kernel of the map 
\begin{equation}
(t_1,t_2) \rightarrow (t_1 t_2^q , t_2^p) \, ,
\end{equation}
which means we can write $t_2 = \zeta_p$ for a primitive $p$-th root of unity $\zeta_p$ and $t_1 = \zeta_p^{-q}$. The action on $\C^2$ with coordinates 
$z_1,z_2$ is hence
\begin{equation}
\label{eq:Crepant}
(z_1,z_2) \rightarrow  (\zeta_p^{-q} z_1, \zeta_p z_2) = (\zeta_p^{p-q} z_1, \zeta_p z_2)  \, .
\end{equation}
It is known that such singularities have a resolution by a chain of rational curves with self-intersection 
numbers $-n_i$ satisfying \cite{Hirzebruch1953}
\begin{equation}
\frac{p}{p-q} = n_1- \frac{1}{n_2 - \frac{1}{n_3 - \frac{1}{ \cdots}}} \, .
\end{equation}
E.g. for $q=1$ with recover the case with crepant resolution for which there are $p-1$ 
curves with $n_i = 2$ appearing in the resolution. Using 
$\frac{1}{2-\frac{n}{n+1}} = \frac{n+1}{n+2}$ then shows the above formula. 

We can use the above formula to deduce the form for some non-crepant singularities that appear frequently in this work. The class we want to consider is a chain that consists of $n+1$ curves with a $(m)$ curve attached to $n\times(2)$ curves. For such cases, we want to find the $\mathbb{Z}_p$ action and the value $q$ in eqn.~\eqref{eq:Crepant}. 
 Setting $n_1=m$ and plugging in the result for the $n$ curves one obtains.
\begin{align}
\label{eq:NonCrepant}
\frac{p}{p-q}=m-\frac{n}{n+1} = \frac{n(m-1)+m}{n+1} \, \quad \text{ with }\,  \left\{ \begin{array}{l}p=n(m-1)+m \\ q= n(m-2) +m-1\end{array}  \right. \, .
\end{align}
This allows to deduce the form of the resolution curves from the values of $q$ mod $p$. Note that we have fixed $q$ to be positive and the $m$ curve to start at the $z_1$ coordinate in the resolution. If $q\neq 1$ we can conjugate $\xi$ by some power to send $q \rightarrow -q$ which can be undone by interchanging $z_1$ and $z_2$ which therefore simply reverses the order of the resolution chain. This allows to deduce the explicit resolution chains that are presented in Section~\ref{ssec:Z5quotient} and~\ref{ssec:Z7quotient}. For $p=5$ we find for $q=3$ just a $(3)(2)$ curve and for $q=2 \sim -3$ the same chain with reversed order. For $p=7$ there is $q=5 $, which is a $(4)(2)$ chain and for $q=4$ this must be a $(3)(2)(2)$. This exhausts all possible values of $q$ for $p=7$.

\section{Compact toric surfaces and elliptic Calabi-Yau threefolds}  
\label{ap:cy3}

In this appendix we review some facts about compact smooth toric varieties and how to construct Calabi-Yau threefolds that are 
elliptically fibered over them. In particular, we will show how this can be done explicitly using reflexive polytopes for all of the 61,539
cases appearing in the classification of \cite{Morrison:2012js}. 

Toric varieties can be described in terms of a fan $\Sigma_B$, and for compact toric surfaces the fan is in turn uniquely determined by giving the ray 
generators $v_i \in \mathbb{Z}^2$. We can label these ray generators such that $i=1..k$ increases when going in a counter-clockwise direction. 
Compactness then implies that the cones of maximal dimensions are spanned by $(v_i,v_{i+1})$, where we have set $v_{k+1}= v_1$. Smoothness now implies 
that the intersections between any of the associated toric divisors $D_i$ are
\begin{equation}
D_i \cdot D_{i+1} = 1 \, . 
\end{equation}
The only other non-zero inner products are the self-intersections $D_i \cdot 
D_i = n_i$. These are determined by the linear relations 
\begin{align}
\sum_i \langle m,v_i \rangle D_i =0 \, \quad \forall m \in \mathbb{Z}^2 \, ,
\end{align}
in particular choosing $m_i$ such that $\langle m_i, v_i \rangle = -1$ implies
\begin{equation}
n_i = \left(\langle m_i, v_{i+1} 
\rangle + \langle m_i, v_{i-1} \rangle  \right) \, .
\end{equation}
Conversely, the ray generators can be uniquely reconstructed from the self 
intersection numbers of toric divisors (up to SL$(2,\Z)$), see 
\cite{Oda1978,Fulton:1436535,Larfors:2010wb} for details.

For a given surface $B$, we may consider the hypersurface 
\begin{equation}\label{eq:tatemodel}
X_0: y^2 + yxw a_1(z) + y w^3 a_3(z) = x^3 + x^2 w^2 a_2(z) + x w^4 a_4(z) + w^6 
a_6(z) \, , 
\end{equation}
in a $\mathbb{P}^2_{123}$ bundle over $B$, and for $a_i(z)$ a holomorphic 
section of $-K_{B}^{\otimes i}$. This space is a Calabi-Yau variety 
which carries an elliptic fibration with base $B$. Under certain 
conditions on the self-intersection numbers of the curves in $B$, this defines 
a sensible compactification of F-Theory to 
6D \cite{Morrison:2012np,Morrison:2012js}. For toric bases, all 61,539 surfaces 
satisfying these conditions have been classified in \cite{Morrison:2012js}.

In the following, we will explain how such a resolution can be explicitly 
found using pairs of reflexive polytopes. For a pair of a polytopes 
$\Delta^*,\Delta$ with vertices in $\mathbb{Z}^4$ satisfying 
\begin{equation}\label{eq:polarduality}
 \langle \Delta,\Delta^* \rangle \geq -1  \, ,
\end{equation}
there exists a smooth family of Calabi-Yau hypersurfaces $X_{\Delta^*,\Delta}$ 
in a toric variety obtained from (a refinement) of the face fan of $\Delta^*$ 
(which is equivalent to the normal fan of $\Delta$) \cite{Batyrev:1994hm}, see 
also \cite{cox1999mirror,Kreuzer:2006ax} for more background on this 
construction. $\Delta$ is then the Newton polytope of the Calabi-Yau hypersurface 
and the topological data of $X_{\Delta^*,\Delta}$ can be computed from 
combinatorial formulas.

The ambient space for the (generally singular) Calabi-Yau $X_0$ can be 
constructed as follows. There is a unique fan in four dimensions with ray 
generators
\begin{equation}\label{eq:naivepilingfan}
v_i = \left(2,3,v_i \right) \, , v_x =  \left(-1,0,0,0 \right)\, , v_y =  \left(0,-1,0,0 \right)\, , v_w =  \left(2,3,0,0 \right)\, ,
\end{equation}
and cones of maximal dimension 
$(\sigma_x,\sigma_y,\sigma_i,\sigma_{i+1}),(\sigma_x,\sigma_w,\sigma_i,\sigma_{
i+1}),(\sigma_w,\sigma_y,\sigma_i,\sigma_{i+1})$.
A anticanonical hypersurface in this toric variety is the given 
by \eqref{eq:tatemodel}. We can construct a crepant 
resolution by performing appropriate blow-ups at the singular loci of the above 
hypersurface.

Due to the simplicity of the toric setup, such resolutions always have a 
description in terms of reflexive polytopes. In the present case, the convex 
hull of the ray generators \eqref{eq:naivepilingfan} defines a
polytope $\Delta^*_0$, which however fails to be reflexive in general. The reason for this is that the polytope $\Delta_0$ defined 
by saturation of the inequality \eqref{eq:polarduality} fails to have vertices at lattice points in $\mathbb{Z}^4$. The geometric reason for 
this is that a generic hypersurface constructed from a pair of reflexive polytopes must be smooth by results of \cite{Batyrev:1994hm}, but the 
Calabi-Yau hypersurface \eqref{eq:tatemodel} is singular. 

The integral points contained in $\Delta_0$, however, precisely correspond to the set of monomials in \eqref{eq:tatemodel}. It is a non-trivial 
fact\footnote{APB would like to thank Washington Taylor and Yi-Nan Wang 
for a collaboration in which this statement was established by scanning through all cases.} that the polytope 
\begin{equation}
\Delta := \mbox{convex hull of} \left\{v | v \in \Delta_0\, \cap\, \Z^4  \right\} 
\end{equation}
is reflexive for all cases in the list of \cite{Morrison:2012js}. The dual $\Delta^*$ of $\Delta$ is hence also reflexive and the pair defines a family of elliptic Calabi-Yau 
hypersurfaces $X_{\Delta^*,\Delta}$. As $\Delta \subseteq \Delta_0$, it follows 
that $\Delta^* \supseteq \Delta^*_0$. As lattice points on $\Delta^*$ correspond 
to 
divisors of $X_{\Delta^*,\Delta}$, this implies that the family 
$X_{\Delta^*,\Delta}$ is a crepant resolution of $X_0$. Furthermore, none of the 
divisors introduced in the resolution 
process is an exceptional divisors of a blow-up purely in the base. This means we can combine the blow-down to $\iota: X \rightarrow X_0$ with the 
projection $\pi_0$ of the elliptic fibration on $X_0$ to find the projection 
$\pi = \pi_0 \circ \iota$ of the elliptic
fibration on $X_{\Delta^*,\Delta}$. We can hence describe a family of smooth 
elliptically fibered Calabi-Yau threefolds by means of a pair of reflexive 
polytopes for any compact 
toric base that leads to a sensible F-Theory compactification to 6D. 
 
The toric description also helps us to discuss the relation between 
the elliptic threefold $X$ over $B$ and the one over $\hat{B}=B/\mathbb{Z}_m$ where $B$ is one of the primaries discussed in Section~\ref{sect:prim_desc}.  First note, that the vertices of the base $\hat{B}$ sit in a lattice of finite index $n$ in $N\sim \mathbb{Z}^2$. This allows us to rewrite the vertices $v_i$ of $\hat{B}$ via the matrix
\begin{align}
M_m \in GL(2,\mathbb{Z}) \qquad  \text{ with } |\text{det}(M_m)|=m \, ,
\end{align} 
that acts as
\begin{align}
M_m \cdot v_i = \hat{v}_i \, .
\end{align}
This point of view is useful as it readily extends to the toric three-fold. E.g. the above toric action can be extended on the toric fan \eqref{eq:naivepilingfan} from which the singular threefold is obtained. This action is simply given as
\begin{align}
 M^\prime_m = \left( \begin{array}{cc}  \mathbf{1}_{2\times2} & 0 \\ 0 & M_m   \end{array} \right) \qquad \text{ with }\quad \text{Det}(\widehat{M})=m \, . 
\end{align} 
which acts on the fan of \eqref{eq:naivepilingfan} via multiplication. 
 Hence the action of $M_m$ extends to a toric action on the ambient space of the toric Calabi-Yau. Now we can also use the Batyrev construction and the algorithm outlined before, to check that this action also extends on the CY hypersurface at special loci in complex structure. To do so, note that that action of $\hat{M}_m$ actions on the polytope of the singular space as
 \begin{align}
 \hat{M}_m \cdot \Delta_0^* = \widehat{\Delta^*}_0 \, .
 \end{align} 
 Then, via the Batyrev construction, there exists a dual polytope $\widehat{\Delta}_0$ that can be obtained from  $\widehat{\Delta^*}_0$ which again consists of the monomials in the CY hypersurface. Since the matrix $M_m$ corresponds to a lattice refinement of $\Delta_0^*$ it reduces those points contained in $\Delta_0$. Hence one may view the quotient action as a complex structure deformation made in $X$ that eliminates all non-quotient invariant monomials.  
 One notices further that the action of $\hat{M}_m$ is trivial on the elliptic fiber direction \footnote{E.g. in \cite{Anderson:2018heq,Anderson:2018kwv,Anderson:2019kmx,Braun:2017feb} examples are given, where the quotient action is extended to the torus fiber as well.} and acts purely in the base. Hence, the Tate coefficients $a_i$ in eqn.~\eqref{eq:tatemodel} must be invariant sections under the quotient actions, which requires the aforementioned complex structure deformation. 
    This is a generalisation to the quotient action of \cite{DelZotto:2014fia} extended to bases  that are not Fano, as well as quotients that lead to $-n>-2$ curves.  
\end{appendix}

\newpage

\providecommand{\href}[2]{#2}\begingroup\raggedright\endgroup

\end{document}